\pdfoutput=1
\documentclass[11pt]{article}
\usepackage{jheppub}
\usepackage{amsthm}
\usepackage{amsmath}
\usepackage{amssymb}
\usepackage{braket}
\usepackage{multirow}
\usepackage{graphicx}
\usepackage{hyperref}
\usepackage{mathrsfs}
\usepackage{subcaption}
\usepackage{empheq}
\usepackage{appendix}
\usepackage{adjustbox}
\usepackage{natbib}
\usepackage[dvipsnames]{xcolor}
\usepackage{float}
\usepackage{slashed}
\usepackage{indentfirst}
\usepackage[top=4.5cm, bottom=-2cm, left=4.5cm, right=-0.5cm, includefoot]{geometry}  
\graphicspath{{Figs/}}
\usepackage{cleveref}
\renewcommand\thefootnote{\textcolor{red}{\fnsymbol{footnote}}}
\def\eq$#1${\begin{equation}#1\end{equation}}
\def\gat$#1${\begin{gather}#1\end{gather}}
\def\gatt$#1${\begin{gather*}#1\end{gather*}}
\def\bal$#1${\begin{align}#1\end{align}}
\def\eqarr$#1${\begin{eqnarray}#1\end{eqnarray}}
\definecolor{nicered}{rgb}{0.7,0.1,0.1}
\definecolor{nicegreen}{rgb}{0.1,0.5,0.1}
\definecolor{rosso}{cmyk}{0,1,1,0.4}
\definecolor{MyDarkBlue}{rgb}{0,0.1,0.7}
\definecolor{PetuniaColor}{RGB}{150,23,147}
\definecolor{myblue}{RGB}{10,10,200}
\definecolor{rossos}{cmyk}{0,1,1,0.55}
\definecolor{green-go}{cmyk}{0.79,0,0.59,0.5}
\definecolor{ForestGreen}{RGB}{34,139,34}
\definecolor{Purple}{RGB}{128,0,128}
\usepackage{hyperref, pdfsync, epstopdf, enumerate}
\hypersetup{colorlinks,bookmarksopen,bookmarksnumbered,
citecolor={nicered},
linkcolor={MyDarkBlue},pdfstartview=FitH,
urlcolor={nicered}}

\newcommand{\newc}{\newcommand}
\newc{\teo}[1]{\textcolor{PetuniaColor}{#1}}
\newc{\corr}[1]{\textcolor{red}{#1}}
\newc{\kostas}[1]{\textcolor{ForestGreen}{#1}}
\newc{\nonum}{\nonumber}
\newc{\pa}{\partial}
\newc{\ra}{\rightarrow}
\newc{\Ra}{\Rightarrow}
\newc{\wtilde}{\widetilde}
\newc{\calG}{\mathcal{G}}
\newc{\calF}{\mathcal{F}}
\newc{\calX}{\mathcal{X}}
\DeclareMathOperator{\sech}{sech}

\makeatletter
\gdef\@fpheader{}
\makeatother
\begin{document}
\title{Fermions and baryons as open-string states from brane junctions}
\author[a]{Theodoros Nakas,}
\author[b]{and Konstantinos S.~Rigatos}
\affiliation[a]{Division of Theoretical Physics, Department of Physics, University of Ioannina, 
Ioannina GR-$45110$, Greece.}
\affiliation[b]{School  of  Physics  \&  Astronomy  and  STAG  Research  Centre,  University  of  
Southampton, Highfield,\\ Southampton  SO$17$ $1$BJ, UK.}
\emailAdd{t.nakas@uoi.gr}
\emailAdd{k.c.rigatos@soton.ac.uk}
\abstract{There has been recent progress towards understanding the dynamics of world-volume fermions 
that arise as open-string modes from brane intersections in the probe limit $(N_f/N_c \rightarrow 0)$. 
In this work we consider all possible BPS brane junctions in Type IIA/B supergravity theories. We study 
in detail the dynamics of these states by deriving their equations of motion. We show the expected 
degeneracy of the bosonic and fermionic fluctuations as is expected due to the preserved supersymmetry. 
We also give some supporting evidence and refine the notion that these states can effectively describe baryon 
operators in a  certain regime of the field theory's parameter space. Our piece of evidence is the demonstration 
of the expected scaling of the mass in the large-$N_c$ limit of the theory for these fermionic states; $M^2 
\sim N^2_c$. Finally, we explain analytically the avoided level crossing that was observed in a previous work after 
the inclusion of higher dimension operators in the field theory.}
\maketitle
\setcounter{page}{1}\setcounter{footnote}{0}
\newpage
\renewcommand*{\thefootnote}{\arabic{footnote}}
\section{Prologue} \label{sec: intro}
The AdS/CFT correspondence \cite{Maldacena:1997re, Witten:1998qj, Gubser:1998bc} asserts that certain field theories are dual (equivalent) to string theory in an AdS space. The original proposal was 
generalized shortly after its discovery to a more general notion of gauge/gravity dual pairs in various dimensions, 
by considering different D$p$-branes \cite{Itzhaki:1998dd}. The prescription is, tersely, to consider a stack 
of $N_c$ D$p$-branes that coincide and take the limit in which the brane modes decouple from the bulk. We 
are left with a super Yang-Mills (SYM) theory with a $U(N_c)$ gauge group on the $(p+1)$-dimensional 
worldvolume of the D$p$ branes. This SYM theory is dual to string theory in the near-horizon limit of the 
background induced by the stack of D$p$-branes. For any D$p$-brane with $p \neq 3$, the holographic gauge 
description is a non-conformal theory, since the Yang-Mills coupling carries dimensions. The energy scale of 
the gauge theory is mapped to the radial coordinate in the gravity side which is orthogonal to the branes. The 
absence of conformal invariance manifests itself in the bulk by the non-trivial radial profile of the dilaton as well as 
the spacetime curvature. The supergravity approximation holds when the string coupling is weak and the 
curvature small. It gives a theory which is trustworthy for an intermediate regime of energies, where the dual 
gauge theory is always strongly coupled. For the case $p=3$, the supergravity description is valid for all energy 
regimes as the dilaton is just a constant. A classic review of the AdS/CFT with a detailed exposition to the above 
as well as other basic ideas is \cite{Aharony:1999ti}. 

In its original form, the AdS/CFT relates Type IIB string theory in $AdS_5 \times S^5$ to the four-dimensional 
$\mathcal{N}=4$ SYM. All matter fields transform in the adjoint representation of the gauge group. For the 
supergravity descriptions of confining gauge theories \cite{Witten:1998zw}, it is possible to calculate the spectra 
of glueball states by considering the corresponding spectra of supergravity modes such that the classical solutions 
to equations of motion are normalisable. There is extensive work on glueball spectra and we cannot do justice to the 
literature so we mention indicatively some of the main papers 
\cite{Csaki:1998qr, Brower:2000rp, Constable:1999gb, Caceres:2005yx}. 

One of the main questions that the community tried to understand since the early days of the duality was how to add fields 
that transform in the fundamental representation of the gauge group, such that the duality contains matter in 
appropriate group representations, and thus taking the duality a step closer to the more interesting quantum field theories 
and thus making the correspondence more physical and appealing. An early variant of the AdS/CFT describing a field 
theory with matter in the fundamental representation related the conformal $Sp(2N_c)$ with $\mathcal{N}=2$ theory 
to D$3$-branes near singularities in F-theory; string theory in $AdS_5 \times S^5 / \mathbb{Z}_2$ 
\cite{Fayyazuddin:1998fb, Aharony:1998xz, Bertolini:2001qa}. 

In a landmark paper \cite{Karch:2002sh}, Karch and Katz showed that adding D$7$ branes in the probe limit---such 
that the D-branes do not backreact to the background geometry---is the appropriate way to add matter fields transforming 
the fundamental representation of the gauge group in the original setup with the D$3$-branes that generate the $AdS_5 
\times S^5$ background. The addition of $N_f$ D$7$ branes in the $AdS_5 \times S^5$ background is equivalent to 
coupling $N_f$ hypermultiplets in the fundamental representation of the original SYM theory. The probe (or in the field 
theory language the quench) limit corresponds to having $N_f/N_c \rightarrow 0$.  The stretched strings between the two different branes are the fields that we consider as dynamical quarks, specifically the lightest string states that stretch between the two different types of branes. Adding the D$7$ branes reduces the supercharges by half compared to the original setup and the resulting four-dimensional $\mathcal{N}=2$ theory has quark-antiquark bound states (mesons) and in the decoupling limit the exercise of studying the dynamics and the masses of these states is equivalent to computing the classical equations of motion of open strings. The seminal paper exploring these ideas is \cite{Kruczenski:2003be} where the analysis was performed for the D$3$-probe D$7$ system. 

A number of directions have been undertaken in order to construct gravity dual descriptions of QCD-like theories. Chiral 
symmetry breaking has been described in \cite{Babington:2003vm, Kruczenski:2003uq, Sakai:2004cn, Klebanov:2000hb} 
and the meson spectra of brane intersections with eight supercharges have been computed in 
\cite{Myers:2006qr, Arean:2006pk}. For more details on flavour physics and mesons in the context of the gauge/gravity 
duality see the review \cite{Erdmenger:2007cm}. The aforementioned works are all in the probe (in gravity terms) or the 
quench limit (in the field theory language) of the correspondence such that the branes do not backreact to the geometry. 
Considerable efforts have been made beyond this approximation. For an excellent review on unquenched flavours in the 
AdS/CFT see \cite{Nunez:2010sf}. 

While much effort has been put and progress has been made towards the understanding of the bosonic sectors of adjoint 
matter fields comprised out of the fundamental quarks (meson fields), much less attention has been given to their 
supersymmetric partners. There are some reasons for this. 

To begin with, in cases with any amount of supersymmetry the spectra of fermionic states can be derived using 
representation theory. This counting had been performed in \cite{Kruczenski:2003be} for the fermionic superpartners of 
the bosonic meson states in the D$3$/probe-D$7$ setup. It is worthwhile mentioning that while it is possible to obtain the 
mass spectra precisely due to the preserved amount of supersymmetry, if one follows this approach then there is no access 
to the dynamics of world-volume fermions as the equations of motion are missing. 

In addition to that, these fermionic superpartners should be formally interpreted as fermionic meson states in the sense 
they are fermionic fields in the adjoint representation of the $\mathcal{N}=2$. Such a state has no counterpart neither in 
the real-world QCD theory nor in any non-supersymmetric version of QCD-like theories. In other words, it can be 
characterized as a purely supersymmetric effect. 

A final hindrance was the form of the fermionic completion to the DBI action. Before the work of 
\cite{Marolf:2003vf, Marolf:2003ye, Martucci:2005rb} that particular part of the action was written in superspace 
\cite{Cederwall:1996ri, Bergshoeff:1996tu}. In spite of the compact form of the action and the elegance of the superspace 
techniques, using the superspace as the target space of the theory obscures how the background fields enter the fermionic
terms of the action. Hence, with that particular formulation of the action at our disposal any explicit computations or considerations 
involving the world-volume fermions cannot be performed. The work of \cite{Marolf:2003vf, Marolf:2003ye, Martucci:2005rb} 
resolved precisely that issue and presented an action for world-volume fermions written in terms of the spacetime. 

Since the technology developed enough and made explicit computations involving world-volume fermions possible the first 
explicit results in probe-brane holography started appearing. This was initiated in \cite{Kirsch:2006he} where the author 
considered the D$3$/probe-D$7$ intersection in the limit where the flavour branes have collapsed on the background 
ones. In field theory terms, this corresponds to massless quarks. From that effective bottom-up approach a replacement 
rule was invoked at the level of the equations of motion that yields the correct mass spectra even for the case of the massive 
embedding of the flavour brane. A follow-up to that work was presented in \cite{Ammon:2010pg} where the authors followed 
the same line of reasoning but for all possible probe-brane setups in the D$3$ background. Another very interesting result was 
derived in \cite{Faraggi:2011bb}, where the authors obtained the spectrum of bosonic and fermionic excitations of $1/2$-BPS 
Wilson loops for a D$3$-brane in the $AdS_5 \times S^5$ background.

With a formal understanding of the massive embedding in the D$3$/probe-D$7$ intersection being a motivation as well as some 
phenomenological applications of holography and the use of fermions in bottom-up models, the authors in \cite{Abt:2019tas} 
studied the top-down massive picture in all probe brane systems in the presence of the D$3$ background and obtained the 
expressions for the supergravity states dual to these fermions and the corresponding mass spectra. The work of \cite{Heise:2007rp} 
is also a top-down approach that investigates the D$4$-D$8$ setup in Type IIA supergravity and it pertains to the Sakai-Sugimoto 
model \cite{Sakai:2004cn}. It is worthwile pointing out that the authors in \cite{Abt:2019tas, Heise:2007rp} are using different 
approaches to finding the mass eigenvalues. In \cite{Abt:2019tas} the authors also considered the effect of adding multi-trace 
operators to these fermionic states following Witten's prescription \cite{Witten:2001ua}. After a careful numerical analysis they 
observed an avoided level-crossing as one approaches from a higher excited state the lower one in the KK tower. 

Another point made by the authors in that paper was that, after analysing the holographic dictionary, they observed that some of 
the aforementioned fermionic states are made of three elementary Fermi fields; two fields transforming in the fundamental 
representation (quarks) and one in the adjoint of the gauge group (a gaugino). Hence, the observation that the lowest lying state 
of that particular supermultiplet resembles qualitatively---by means of the same structure---the baryon operators in real-life QCD.

In this work we continue in a natural line of the aforementioned research and we study all possible probe-branes intersections 
in Type IIA/B supergravity theories that have been shown to preserve eight supercharges; $\mathcal{N}=2$ SUSY in a 
four-dimensional language. Since the bosonic mass spectra are explicitly known in these cases \cite{Myers:2006qr, Arean:2006pk} 
the interest lies in the deeper understanding of the dynamics of fermionic open-string states and unveiling potential computational 
subtleties in the derivation that did not appear in the study of probe-branes in the presence of the D$3$-background. We obtain 
the equations of motion explicitly for each case starting from the fermionic action as was derived in 
\cite{Marolf:2003vf, Marolf:2003ye, Martucci:2005rb} . 

We show the expected supersymmetric degeneracy for the mass eigenvalues. For the case of the D$3$-background there are analytic 
solutions that have been previously obtained \cite{Abt:2019tas} and we report these here for completeness. All the other backgrounds 
do not admit analytic solutions for the eigenstates of the equations of motion and the mass eigenvalues. For these cases we work in two 
different ways. To begin with, we perform a careful numerical analysis and obtain the mass eigenvalues directly from the fermionic 
equations of motion. We see that in each case they are the same as the corresponding results anticipated from the analysis of the bosonic 
meson states and their equations of motion.

Furthermore, we were able to derive a map such that we bring the equations of motion describing world-volume fermions in our 
brane intersections to the relevant equations of motion describing bosonic degrees of freedom in these setups. Here some comments 
are in order. Due to the two eigenvalues of the spinor spherical harmonics we have two sets of supergravity fields. We denote those 
associated with the positive eigenvalues by $\mathcal{G}$. These are the superpartners of the gauge fields on the probe-brane which 
have a non-trivial profile only on the internal manifold. More explicitly, these gauge fields are decomposed as $A_{\mu}= 0, \quad 
A_{\rho} = 0, \quad A_{i} = a(\rho) e^{i k^{\mu} x_{\mu}} \mathcal{Y}^{\ell^\pm}_{i}$. From the negative eigenvalue of the spinor 
spherical harmonic we obtain modes which we generically denote by $\mathcal{F}$ and they are the superpartners of the scalar fields 
obtained as fluctuations of the transverse coordinates.

For both types of the fermionic modes described above we were able to map the equations of motion associated with the positive eigenvalue 
of the projection $\Gamma$-matrix defined along the holographic radial coordinate ($\Gamma^\rho\Psi^{\pm}=\pm\Psi^{\pm}$) to the 
associated bosonic equations of motion. The map is given by
\begin{equation} \label{eq: basic_1}
\psi^{\oplus}_{\mathcal{G}_\ell}(\varrho) = \frac{1}{(1+\varrho^2)^{\frac{3}{16}(7-p)}} f_{\ell+1}(\varrho), \qquad 
\psi^{\oplus}_{\mathcal{F}_\ell}(\varrho) = \frac{1}{(1+\varrho^2)^{\frac{3}{16}(7-p)}} f_{\ell}(\varrho),
\end{equation}
where in the above we denote by $\oplus$ superscript the supergravity modes obtained by considering the projection $\Gamma^\rho
\Psi^{\oplus}=\Psi^{\oplus}$ and we reserve the $\ominus$ for the negative eigenvalue of the projection.

While at this point the reader might be concerned that we were able to map only the equations of motion related to one 
of the $\Gamma^\rho$ eigenvalues, there is nothing out of place in doing so. We are able to deduce the eigenstates and mass 
eigenvalues from either of the equations of motion. The reasoning behind this is simple: we start with two coupled first-order equations 
of motion that describe a source and the relevant operator. Then, we derive the second-order equations of motion starting from the 
first-order ones. In doing so we act with an appropriately chosen differential operator and we create a double-copy of the operator 
and the source \cite{Laia:2011wf, Abt:2019tas}. Hence, the equations of each eigenvalue of the $\Gamma^\rho$ matrix contains a 
copy of the source and a copy of the operator. By requiring normalizable open-string modes in the UV in either of the equations we are 
setting the source to zero and hence each second-order differential equation yields the correct answer individually. This procedure 
is made more precise and explained thoroughly in \Cref{sec: D0}. We also, further support and refine the argument that these states 
describe effectively baryons as we show that in large-$N_c$ limit their mass scales as 
\begin{equation} \label{eq: basic_2}
M^2 \sim N^2_c\,,
\end{equation}
when we fix the quantum number $n$ which counts the nodes of the state and the angular quantum number $\ell$. This scaling in 
large-$N_c$ limit of the theory is a purely field theory expectation \cite{Witten:1979kh}. The results described by \cref{eq: basic_1,eq: basic_2} 
constitute the two main points of this work. The third main result is the explanation of the avoided level crossing, which is explained in 
\Cref{sec: multi_trace}. 

Furthermore, the authors of \cite{Abt:2019tas} observed an avoided level crossing when examining the inclusion of multi-trace 
interactions without giving an analytic explanation for the effect. We revisit that computation and give an analytic justification to 
the observed avoided level crossing by bringing the equations of motion in a Schr\"odinger form and examining the solutions of the 
supergravity modes. Interestingly, such a crossing has been shown to occur when considering instanton configurations in the 
probe-brane setup \cite{Erdmenger:2005bj}. 

At this point, we would like to clarify that we do not claim that we have solved the true string baryonic vertex. In the 
large-$N_c$ limit a baryon is made of $N_c$ fundamental fields (quarks) and has to be appropriately described in the gravity 
side. The description is well known \cite{Witten:1998xy}. Witten has shown that in the base $\mathcal{N}=4$ SYM theory a baryonic 
vertex is obtained by a wrapped D$5$-brane over the $S^5$. Since we have not considered the addition of the five-branes in our 
basic D$3$/probe-D$7$ model, we do not approach baryons directly in this work. However, we believe that the spectrum we obtain 
in \Cref{sec: D3_largeNc} should be obtained from the full string construction in the limit where the boundary gauge theory is in 
the conformal window. This result was obtained in an effective, bottom-up way in  \cite{Kirsch:2006he}, however a $10$-dimensional 
derivation was lacking, which we provide here.

This is a long and technical paper which we organized in such a way to help the reader navigate more easily through the different 
sections. The structure of the paper is as follows: in \Cref{sec: generalities} we present the geometries that are related to our studies 
and compute some necessary quantities for our computations. We devote a separate section for each supergravity background. The 
D$0$-background is analysed very thoroughly in \Cref{sec: D0} and it serves as a specific example which we use to demonstrate 
the general steps of our analysis. After this, we recap the steps necessary and explain the general procedure we are following in 
\Cref{sec: general idea}. Then, \Cref{sec: D1} is devoted to the D$1$ solution of IIB. The case of the D$2$-background branes is 
considered in \Cref{sec: D2}. Probe flavour branes in the D$3$ solution are considered in \Cref{sec: D3} and the D$4$-background 
is presented in \Cref{sec: D4}. We conclude and discuss future directions in \Cref{sec: outro}. We include some appendices that 
supplement the material of the main body of the work. In \Cref{app: conventions} we explain our notation and our conventions. In 
\Cref{app: unitsphere} we present some explicit geometrical results related to the unit $N$-sphere. Finally, in \Cref{app: numerical} 
we present very briefly the equations of motion for the scalar mesons in the probe-brane setups under consideration and state the 
tuned parameters that are necessary to obtain the mass eigenvalues in the cases that do not admit analytic solutions. 
\newpage
\section{Background spacetime, probe-brane geometries, and the general approach} \label{sec: generalities}
In this section we wish to establish notation and conventions, as well as set the stage for the analysis of the forthcoming sections. 
The analysis presented below in \cref{sec: thedpbrane} and \cref{sec: inducedgeometries} has been performed a number of times in 
the past \cite{Myers:2006qr, Arean:2006pk, Johnson:2000ch}, however we do find it convenient and useful to quote once more the 
basic results here. 
\subsection{The background geometry}\label{sec: thedpbrane}
We will be considering the brane junctions of a D$p$-brane and a D$k$-brane (under the assumption that $p \leq k$) along a 
number $d$ of common spatial dimensions such that the system is $1/2$-BPS. We will denote this intersection by $\{d,p,k \}$. 
The lower dimensional brane is treated as the background one, while we treat the other one as the probe. Taking into consideration 
the above physical constraints, one can verify that there are three possibilities for consistent intersections: $\{p,p,p+4 \}$, 
$\{p-1,p,p+2 \}$, and $\{p-2,p,p \} $.

The supergravity background geometry that describes a stack of $N_c$ coincident D$p$-branes is given by
\begin{align} \label{eq: supergr-geo}
ds^2 = \frac{1}{\sqrt{H_{p}}} dx^{\mu} dx_{\mu} + {\sqrt{H_{p}}} ~ d \vec{Z} \cdot d \vec{Z}\,,
\end{align}
written in the string frame. The $\mu$ is taking values over a ${\left(p+1\right)}$-dimensional Minkowski spacetime parallel to the 
branes, while the $Z$-coordinates parametrize the $\left( 9 -p \right)$-dimensional transverse space. In the above, $p \leq 4$, 
while the expression of the harmonic function $H_p$ is of the following form 
\eq$
H_{p} = 1+ \left(\frac{R}{r}\right)^{7-p},$
with $R$ being given by
\begin{align}
R^{7-p} = 2^{5-p} \pi^{(5-p)/2}~ g_{s} ~ N_c ~ \Gamma \left( \frac{7-p}{2}  \right) (\alpha')^{(7-p)/2},
\end{align}
where in the above $N_c$ is the number of the background D$p$-branes, $g_s$ is the string coupling constant, and $\alpha'$ is 
the inverse string tension. We will be working in the decoupling limit where the open string modes decouple from the closed string 
modes, thus, $\alpha'\ra 0$ and
\eq$
\lim_{\alpha'\ra 0}g^{2}_{\textit{YM}}=\lim_{\alpha'\ra 0}(2 \pi)^{p-2} ~ g_{s} ~ {\alpha'}^{(p-3)/2}=\textit{fixed}\,.$

Taking the near-horizon limit, the geometry of \cref{eq: supergr-geo} takes the form
\begin{align}
ds^2 &= \left( \frac{r}{R} \right)^{\frac{7-p}{2}} dx^{\mu} dx_{\mu} + \left( \frac{R}{r} \right)^{\frac{7-p}{2}} ~ d \vec{Z} \cdot 
d \vec{Z}, 
\end{align}
while the dilaton and the Ramond-Ramond (R-R) potential can be expressed in terms of the $r$-coordinate as follows
\begin{align}\label{eq: dil&RRpot}
e^{\phi} &= \left(\frac{R}{r}\right)^{\frac{\left(7-p \right)\left(3-p \right) }{4}}, &&& C_{(p+1)}=\left( \frac{r}{R}\right)^{7-p}\, 
dx^0\wedge\ldots\wedge dx^p\,.
\end{align}

For probe embeddings a particular useful re-parametrization of the near-horizon geometry is to choose spherical coordinates 
and re-express the spacetime. This can be achieved as follows: consider the introduction of the probe D$k$-brane in this 
background which extends along the directions 
\begin{equation}
X^{\hat{A}} = \{t, x^1, x^2, \cdots, x^d, Y^1, Y^2, \cdots, Y^{k-d} \}.
\end{equation}
The hatted capital Latin indices denote the coordinates of the probe D$k$-brane. It is worthwhile stressing that the set of 
$Y$-coordinates describes directions that are transverse to the background D$p$-brane, which means that they coincide 
with some of the $Z$-coordinates. The rest of the $Z$-coordinates which are transverse to both background and probe 
branes will be denoted with $\vec{w}$, namely
\begin{equation}
\vec{w} = \{w^1, w^2, \cdots, w^{9-p-k+d} \},
\end{equation}
with $w^{m} = Z^{k-d+m}$, and $m=1,2,\cdots,9-p-k+d$.  In order to introduce spherical coordinates on the world-volume 
of the D$k$-brane that is transverse to the D$p$-brane, we define
\bal$\label{eq: sphercord-Dk}
&Y^1=\rho\, \cos\theta_1\,,\nonum\\
&Y^2=\rho\, \sin\theta_1\,\cos\theta_2\,,\nonum\\
&\hspace{0.25em}\vdots \\
&Y^{k-d-1}=\rho\,\sin\theta_1\,\sin\theta_2\cdots\sin\theta_{k-d-2}\,\cos\theta_{k-d-1}\,,\nonum\\
&Y^{k-d}=\rho\,\sin\theta_1\,\sin\theta_2\cdots\sin\theta_{k-d-2}\,\sin\theta_{k-d-1}\,,\nonum$
which results to
\begin{align}
\left( {Y^{1}} \right)^{2} + \left( {Y^{2}} \right)^{2} + \cdots + \left({Y^{k-d}} \right)^2=\rho^2\,,
\end{align}
and
\begin{equation}
\left( {d Y^{1}} \right)^{2} + \left( {d Y^{2}} \right)^{2} + \cdots + \left({d Y^{k-d}} \right)^2=d \rho^2 + \rho^2 
d \Omega^2_{k-d-1}\,.
\end{equation}
In the above relation
\eq$d \Omega_{k-d-1}^2=d\theta_1^2+\sum_{i=2}^{k-d-1}\left(\prod_{j=1}^{i-1}\sin^2\theta_j\right)d\theta_i^2\,,$ 
is the line element of the unit $\left( k-d-1 \right)$-dimensional sphere.

Consequently, decomposing the $Z$-coordinates in the $Y$ and $w$-coordinates, which are parallel and transverse to the 
D$k$-brane, respectively, we are led to
\begin{align} \label{eq: 10d_geometry}
ds^2 = \left(\frac{r}{R}\right)^{\frac{7-p}{2}}  \sum^{p}_{\mu,\nu=0}\eta_{\mu \nu}\,dx^{\mu}\, dx^{\nu} &+ 
\left( {\frac{R}{r}}\right)^{\frac{7-p}{2}} \left(d \rho^2 + \rho^2\, d \Omega_{k-d-1}^2+\sum^{9-p-k+d}_{m,n=1}
\delta_{mn}\,dw^{m}\,dw^{n}\right)\,,
\end{align}
where $r^2=\rho^2+\sum_{m=1}^{9-p-k+d}(w^m)^2$.
\subsection{The geometries on the probe branes} \label{sec: inducedgeometries}
Let us start by considering the embedding of D$\left(p+4\right)$ branes, with $0 \leq p \leq 4$. We embed the flavour 
probe-brane in the $x^{p+5}x^{p+6}\cdots x^9$-plane at a constant position $|\vec{w}|=L\neq0$. The induced geometry on 
the probe can be written as 
\begin{align}
ds^2 = \left( \frac{r}{R} \right)^{\frac{7-p}{2}} ds^2\left( \mathbb{M}^{(1,p)} \right) + \left( \frac{R}{r} \right)^{\frac{7-p}{2}} 
d \rho^2  + \left( \frac{R}{r} \right)^{\frac{7-p}{2}} \rho^2\, d \Omega_{3}^2\,,
\end{align}
where $\mathbb{M}^{(1,p)}$ denotes the $(1+p)-$dimensional Minkowski spacetime. The boundary gauge theory is a 
$(p + 1)$-dimensional SYM theory coupled to a matter hypermultiplet in the fundamental representation of the gauge group and 
the quark mass is proportional to the distance $L$ that separates the two different branes. In case of $p=3$ the induced 
geometry in the UV limit ($r=\rho\ra\infty$) spans an $AdS_5\times S^3$ spacetime.

Now, we want to consider the supersymmetric embedding of D$\left(p+2\right)$ branes, with $1 \leq p \leq 4$. In this case 
the dual gauge theory is $(p + 1)$-dimensional with the fundamental hypermultiplet confined on a $p$-dimensional surface. 
The induced geometry is equal to 
\eq$
ds^2 = \left( \frac{r}{R} \right)^{\frac{7-p}{2}} ds^2\left( \mathbb{M}^{(1,p-1)} \right) + \left( \frac{R}{r} \right)^{\frac{7-p}{2}} 
d \rho^2  + \left( \frac{R}{r} \right)^{\frac{7-p}{2}}  \rho^2\, d \Omega_{2}^2\,.$
In case of $p=3$ the induced geometry in the UV limit spans an $AdS_4\times S^2$ geometry.

Finally, we want to consider embedding probe D$p$-branes in the background of geometry of D$p$-branes with $2 \leq p \leq 4$. 
The matter field of fundamental hypermultiplet propagate in a co-dimension two defect; the hypermultiplet is confined on a 
$\left(p-1\right)$-dimensional surface. The geometry on the probe brane can be written as
\eq$
ds^2 = \left( \frac{r}{R} \right)^{\frac{7-p}{2}} ds^2\left( \mathbb{M}^{(1,p-2)} \right)  + \left( \frac{R}{r} \right)^{\frac{7-p}{2}} 
d \rho^2  + \left( \frac{R}{r} \right)^{\frac{7-p}{2}} \rho^2\, d \Omega_{1}^2. $
In case of $p=3$ the induced geometry in the UV limit spans an $AdS_3\times S^1$ spacetime.
\subsection{Vielbeins  \& spin-connection components} \label{sec: generalformulae}
An appropriate choice of basis will be proven very helpful for the forthcoming calculations, thus, using the vielbein formalism---for 
which it holds that $g_{MN} = e^{(A)}{}_M \, e^{(B)}{}_N \, \eta_{(A)(B)}$\footnote{Indices between brackets will be used solely to 
denote vielbein indices.}---in the geometry of \cref{eq: 10d_geometry}, it is straightforward to evaluate the following 
quantities:
\bal$ \label{eq: vielbeine-1}
&e^{(A)}{}_{\mu}= \left( \frac{r}{R} \right)^{\frac{7-p}{4}} \delta^A{}_{\mu}, \hspace{2em} e^{(A)}{}_{\rho} = \left( \frac{R}{r} 
\right)^{\frac{7-p}{4}} \delta^A{}_{\rho}, \\[2mm]
\label{eq: vielbeine-2}
&e^{(A)}{}_{\bar{i}}=\rho \left( \frac{R}{r} \right)^{\frac{7-p}{4}}\,\bar{e}^{(A)}{}_{\bar{i}}\,,\quad i =1, \ldots, 
k-d-1\,,\\[2mm]
\label{eq: vielbeine-3} 
&e^{(A)}{}_{\wtilde{m}} = \left( \frac{R}{r} \right)^{\frac{7-p}{4}} \delta^A{}_{\wtilde{m}}\,, \quad m=1,\ldots,9-p-k+d\,, $
where the indices with bar correspond to the spherical coordinates $\{\theta_1,\ldots,\theta_{k-d-1}\}$, while the indices with tilde 
correspond to the coordinates $\{w^1,\ldots,w^{9-p-k+d}\}$. In addition, the quantity $\bar{e}^{(A)}{}_{\bar{i}}$ represents the 
vielbein of $S^{k-d-1}$; its explicit expression is given in \Cref{app: unitsphere}. The vielbein basis  $\left\{e^{(A)}\,\big|\,e^{(A)}
=e^{(A)}{}_M\,dx^M\right\}$\footnote{The dual basis $\left\{e_{(A)}\,\big|\, e_{(A)}=e_{(A)}{}^M\,\pa_M\right\}$ is defined through 
the relation \eq$e^{(A)}\left(e_{(B)}\right)=e^{(A)}{}_M\,dx^M \left(e_{(B)}{}^N\,\pa_N\right)=e^{(A)}{}_M\,e_{(B)}{}^N\,\delta^M{}_N
=\delta^{(A)}{}_{(B)}\,.$}, satisfy the following relation
\eq$
ds^2=g_{MN}\,dx^M\,dx^N=\eta_{(A)(B)}\,e^{(A)}\,e^{(B)}\,.$
We also present below the non-zero components of the spin-connection. We remind the readers that for torsion-free theories the 
definition is
\begin{equation} \label{eq: spin_connection_def}
\omega^{(A)(B)}_M = e^{(A)}{}_N \, \Gamma^N_{ML}\,e^{(B)L} + e^{(A)}{}_N \, \partial_M \, e^{(B)N},
\end{equation}
where $e^{(A)M}=e^{(A)}{}_N\,g^{NM}=\eta^{(A)(B)}e_{(B)}{}^M$. Using \cref{eq: spin_connection_def} we obtain the following 
spin-connection components:
\bal$\label{eq: spinconnection-1}
\omega_\mu^{(A)(B)}	&= \frac{7-p}{4}\frac{r^{\frac{3-p}{2}}}{R^{\frac{7-p}{2}}} \left[\rho\big(\delta^A_\mu\,\delta^B_\rho-
\delta^B_\mu\,	\delta^A_\rho\big)+w^{\wtilde{m}}	\big(\delta^A_\mu\,\delta^B_{\wtilde{m}}-\delta^B_\mu\,\delta^A_{\wtilde{m}}
\big) \right]\,,\\[2mm]
\label{eq: spinconnection-2}
\omega^{(A)(B)}_\rho&=\frac{(7-p)}{4} \frac{w^{\wtilde{m}}}{r^2}\big(\delta^A_{\wtilde{m}}\,\delta^B_\rho-\delta^B_{\wtilde{m}}\,
\delta^A_\rho\big)\,,\\[2mm]
\label{eq: spinconnection-3}
\omega^{(A)(B)}_{\bar{k}}	&=\bar{\omega}^{(A)(B)}_{\bar{k}}+\left[\frac{(7-p)\rho^2}{4r^2}-	1\right]\big(	\delta^A_\rho\,
\bar{e}^{(B)}{}_{\bar{k}}-	\delta^B_\rho\,\bar{e}^{(A)}{}_{\bar{k}}\big)+\frac{(7-p)\rho\, w^{\wtilde{m}}}{4 r^2}\big(
\delta^A_{\wtilde{m}}\,\bar{e}^{(B)}{}_{\bar{k}}-\delta^B_{\wtilde{m}}\,	\bar{e}^{(A)}{}_{\bar{k}}\big)\,,\\[2mm]
\label{eq: spinconnection-4}
\omega^{(A)(B)}_{\wtilde{m}}&=\frac{(7-p)}{4 r^2} \left[\rho\big(\delta^A_\rho\,\delta^B_{\wtilde{m}}-\delta^B_\rho\,
\delta^A_{\wtilde{m}}	\big)+ w^{\widetilde{n}}\big(\delta^A_{\wtilde{n}}\,\delta^B_{\wtilde{m}}-\delta^B_{\wtilde{n}}\,
\delta^A_{\wtilde{m}}\big) \right]\,,$
where the quantity $\bar{\omega}^{(A)(B)}_{\bar{k}}$ constitutes the spin-connection of $S^{k-d-1}$; its explicit expression is 
given in \Cref{app: unitsphere} as well. 
\subsection{The Dirac operator} \label{sec: dirac_operator}
The Dirac operator for fermionic fields on a generic curved manifold is defined by 
\eq$\label{eq: diracoperatorone}
\slashed{D} = \Gamma^{N}\,e_{(N)}{}^{M}\left( \partial_{M} + \frac{1}{8}\,\omega^{(K)(L)}_{M}\,[\Gamma_{K},\Gamma_{L}]\right)=
\Gamma^{N}\,e_{(N)}{}^{M}\left( \partial_{M} + \frac{1}{4}\,\omega^{(K)(L)}_{M}\,\Gamma_{KL}\right)\,.$
In what follows, we compute the Dirac operator for the general case that we consider here; the embedding of a probe D$k$-brane 
in the background generated by D$p$-branes. Note that for a probe D$k$ the index $M$ in \cref{eq: diracoperatorone} should be 
replaced by $\hat{M}$, which runs over the probe brane coordinates. After a straightforward computation we are led to
\bal$\label{eq: Dkdirac}
\slashed{D}_{Dk}=&\frac{R^{\frac{7-p}{4}}}{r^{\frac{7-p}{4}}}\,\Gamma^{\mu}  \partial_{\mu} + \frac{r^{\frac{7-p}{4}}}{R^{\frac{7-p}{4}}}\, 
\Gamma^{\rho}  \partial_{\rho} +\frac{1}{\rho}\frac{r^{\frac{7-p}{4}}}
{R^{\frac{7-p}{4}}} \slashed D_{S^{k-d-1}} +\frac{1}{R^{\frac{7-p}{4}}}\left[ \frac{7-p}{8} \frac{\rho}{r^{\frac{p+1}{4}}}(2d+2-k)+\frac{1}{2} 
\frac{r^{\frac{7-p}{4}}}{\rho} (k-d-1)\right]  \Gamma^{\rho} \nonum\\
&+\frac{7-p}{8} \frac{1}{R^{\frac{7-p}{4}}} \frac{1}{r^{\frac{p+1}{4}}} \left(2d + 1 - k \right) \sum^{9-p-k+d}_{m=1} w^{m}\,
\Gamma_{\widetilde{m}}\,,$
with
\eq$\label{sph-dir}
\slashed D_{S^{k-d-1}}=\Gamma^{N}\,\bar{e}_{(N)}{}^{\bar{k}}\left( \partial_{\bar{k}} + \frac{1}{4}\bar{\omega}^{(K)(L)}_{\,\bar{k}}\,
\Gamma_{KL}\right).$
We have already mentioned that the consistent intersections---characterized by the set of numbers $\{d,p,k\}$---between the background D$p$- 
and probe D$k$-branes, are: $\{p,p,p+4 \}$, $\{p-1,p,p+2 \}$, and $\{p-2,p,p \} $. One can easily verify that in all cases $2d+1-k=p-3$, thus, 
\cref{eq: Dkdirac} can be written as
\bal$\label{eq: Dkdirac-sec}
\slashed{D}_{Dk}=&\frac{R^{\frac{7-p}{4}}}{r^{\frac{7-p}{4}}}\,\Gamma^{\mu}  \partial_{\mu} + \frac{r^{\frac{7-p}{4}}}{R^{\frac{7-p}{4}}}\, 
\Gamma^{\rho}  \partial_{\rho} +\frac{1}{\rho}\frac{r^{\frac{7-p}{4}}}
{R^{\frac{7-p}{4}}} \slashed D_{S^{k-d-1}} +\frac{1}{R^{\frac{7-p}{4}}}\left[ \frac{(7-p)(p-2)}{8} \frac{\rho}{r^{\frac{p+1}{4}}}+\frac{1}{2} 
\frac{r^{\frac{7-p}{4}}}{\rho} (k-d-1)\right]  \Gamma^{\rho} \nonum\\
&-\frac{(7-p)(3-p)}{8} \frac{1}{R^{\frac{7-p}{4}}} \frac{1}{r^{\frac{p+1}{4}}} \sum^{9-p-k+d}_{m=1} w^{m}\,
\Gamma_{\widetilde{m}}\,.$
\section{D0 branes: supersymmetric matrix quantum mechanics} \label{sec: D0}

In the background generated by a stack of D$0$-branes there is a unique way to arrange the flavour branes which we demonstrate in 
\Cref{table: D0_brane_intersections}. 
\begin{table}[H]
\begin{center}
\begin{tabular}{ |c|c|c|c|c|c|c|c|c|c|c|c|}
 \hline
 &&&&&&&&&&\\[-0.95em] 
   									& $x^0$ & $x^1$ & $x^2$ & $x^3$ & $x^4$ & $x^5$ & $x^6$ & $x^7$ & $x^8$ & $x^9$			\\ 
 \hline
 Background D0-brane 	& --- & $\bullet$ & $\bullet$ & $\bullet$ & $\bullet$ & $\bullet$ & $\bullet$ & $\bullet$ & $\bullet$ & $\bullet$	\\ 
 \hline
 Probe D4-brane 			& --- & --- & --- & --- & --- & $\bullet$ & $\bullet$ & $\bullet$ & $\bullet$ & $\bullet$				\\
 \hline
\end{tabular}
\caption{The brane intersection. In the above notation --- denotes that a brane extends along that particular direction, while $\bullet$ means 
that the coordinate is transverse to the brane.}
\label{table: D0_brane_intersections}
\end{center}
\end{table}
The fermionic action \cite{Marolf:2003vf, Marolf:2003ye, Martucci:2005rb} is given by
\begin{equation}\label{eq: D0D4action}
S_{D4} = \frac{T_{D4}}{2} \int d^{5} \xi\, \sqrt{-\hat{g}}\, \bar{\Psi}\, \mathcal{P}_{-} \left[ \slashed{D}_{D4}
-\frac{e^{\phi}}{8\cdot 2!}F_{AB}\left(\Gamma^{\hat{M}}\Gamma^{AB}\Gamma^{(10)}\Gamma_{\hat{M}}+3\,\Gamma^{AB}\Gamma^{(10)}\right)
-\frac{1}{2} \Gamma^M \partial_M \phi \right] \Psi
\end{equation}
In the above, $(T_{Dk})^{-1}=(2\pi)^k(a')^{\frac{k+1}{2}}\,g_s$ is the brane tension, $\mathcal{P}_{-}$ is a $\kappa$-symmetry projector ensuring
$\kappa$-symmetry invariance of the action, and $F_{AB}$ represents the components of the 2-form R-R field strength. With 
the use of \cref{eq: dil&RRpot} and setting $p=0$ it is straightforward to determine both the dilaton field 
\eq$\label{eq: dil04}
e^{\phi}=\left(\frac{R}{r}\right)^{\frac{21}{4}},$
and the 2-form R-R field strength
\eq$\label{eq: RR04}
F_{(2)}=dC_{(1)}=-\frac{7r^5}{R^7}\, e^{(0)}\wedge\left(\rho\, e^{(\rho)}+\sum_{i=1}^5 w^i\,
e^{(w^{i})}\right)\,.$
In addition, $\Gamma^{(10)}\equiv \Gamma^{01\cdots p\,\rho\, \theta_1\cdots \theta_{k-d-1}\,w^1\cdots w^{9-p-k+d}}$ is the ten-dimensional 
chiral operator, $\hat{g}_{\hat{M}\hat{N}}=P[g]_{\hat{M}\hat{N}}$ is the pullback of the background metric $g_{MN}$ on the worldvolume, 
while $\hat{g}$ constitutes the determinant of the aforementioned metric. 

Varying \cref{eq: D0D4action} with respect to the conjugate spinor  $\bar\Psi$, we readily obtain the equation of motion of the spinor $\Psi$ 
($\Psi$ is a ten-dimensional spinor of positive chirality $\Gamma^{(10)}\Psi=\Psi$), namely
\eq$\label{eq: D0D4eqmot}
 \slashed{D}_{D4}\Psi-\frac{e^{\phi}}{8\cdot 2!}F_{AB}\left(\Gamma^{\hat{M}}\Gamma^{AB}\Gamma^{(10)}\Gamma_{\hat{M}}+3\Gamma^{AB}
 \Gamma^{(10)}\right)\Psi-\frac{1}{2} (\Gamma^M \partial_M \phi)\Psi=0\,.$
 Using eqs. \eqref{eq: dil04} and \eqref{eq: RR04} together with the properties of $\Gamma^M$ matrices, we can evaluate
\eq$\label{eq: D0D4gammaterm}
\frac{e^{\phi}}{8\cdot 2!}F_{AB}\left(\Gamma^{\hat{M}}\Gamma^{AB}\Gamma^{(10)}\Gamma_{\hat{M}}+3\Gamma^{AB}\Gamma^{(10)}\right)\Psi=
-\frac{7}{4R^{7/4}}\frac{\rho}{r^{1/4}} \,\Gamma^{0\rho}\Psi=\frac{7}{4R^{7/4}}\frac{\rho}{r^{1/4}} \,\Gamma_{0\rho}\Psi\,,$
while by setting $p=d=0$ and $k=4$ in \cref{eq: Dkdirac} we are led to
\bal$\label{eq: D0D4dirac}
\slashed{D}_{D4}\Psi=&\left\{\left(\frac{R}{r}\right)^{\frac{7}{4}}\Gamma^\mu\pa_\mu+\left(\frac{r}{R}\right)^{\frac{7}{4}}\Gamma^\rho
\pa_\rho+\frac{1}{\rho}\left(\frac{r}{R}\right)^{\frac{7}{4}}\slashed{D}_{S^3}+
\frac{1}{R^{\frac{7}{4}}}\left(-\frac{7}{4}\frac{\rho}{r^{\frac{1}{4}}}+\frac{3r^{\frac{7}{4}}}{2\rho}\right)\Gamma^\rho\right.\nonum\\[2mm]
&\left.-\frac{21}{8}\frac{1}{R^{\frac{7}{4}}}\frac{1}{r^{\frac{1}{4}}}\sum_{i=1}^{5}w^i\,\Gamma_{w^i}\right\}\Psi\,.$
For the spherical piece we can utilize the spinor spherical harmonics \cite{Camporesi:1995fb}
\begin{equation} \label{eq: D0D4harm}
\slashed{D}_{S^{3}}\Psi^{\ell^{\pm}} = \pm \left(\ell + \frac{3}{2} \right)\Psi^{\ell^{\pm}}\,.
\end{equation}
It is also important to notice that the above calculations have been performed in the vielbein basis $\{e^{(M)}\}$ (see for example the expression 
of the R-R field strength) instead of the usual $\{dx^M\}$. Thus, we should recognise that $\Gamma^M\pa_M\phi$ is in fact $\Gamma^N 
e_{(N)}{}^M\pa_{M}\phi$. From \cref{eq: dil&RRpot} and after a straightforward calculation, we get
\eq$\label{eq: D0D4dilterm}
\Gamma^M\pa_M\phi\ra\Gamma^N e_{(N)}{}^M\pa_{M}\phi=-\frac{21}{4}\frac{1}{R^{\frac{7}{4}}}\frac{1}{r^{\frac{1}{4}}}\left(\rho\,\Gamma^\rho+
\sum_{i=1}^{5}w^i\,\Gamma_{w^i}\right)\,.$

Consequently, substituting  eqs. \eqref{eq: D0D4gammaterm}-\eqref{eq: D0D4dilterm} in \cref{eq: D0D4eqmot}  and imposing the projection 
$\Gamma_{0\rho}\Psi=-\Psi$, we are led to the following first-order equation of motion
\begin{equation} \label{eq: first_order_D0D4}
\left\{ \left( \frac{R}{r} \right)^{\frac{7}{4}} \Gamma^{\mu} \partial_{\mu} +  \left( \frac{r}{R} \right)^{\frac{7}{4}} \Gamma^{\rho} \partial_{\rho}
+\frac{1}{R^{\frac{7}{4}}} \left[\frac{7}{4} \frac{\rho}{r^{\frac{1}{4}}} \pm \left( \ell + \frac{3}{2}  \right) \frac{r^{\frac{7}{4}}}{\rho} \right] 
+\frac{1}{R^{\frac{7}{4}}}\left( \frac{7}{8} \frac{\rho}{r^{\frac{1}{4}}} + \frac{3}{2} \frac{r^{\frac{7}{4}}}{\rho} \right) \Gamma^{\rho}\right\}
\Psi^{\ell^{\pm}} = 0\,.
\end{equation}

In what follows we consider that $\sum_{i=1}^5 (w^i)^2=L^2$, which by its turn leads to $r^2=\rho^2+L^2$. $L$ constitutes the distance between 
the D$0$- and the D$4$-branes in the $x^5 x^6 x^7 x^8 x^9$-hyperplane of the 10-dimensional spacetime. Acting with the operator $\left( 
\frac{R}{r}\right)^{7/4} \Gamma^{\mu} \partial_{\mu} +\left( \frac{r}{R} \right)^{7/4} \Gamma^{\rho} \partial_{\rho}$ on the l.h.s. of 
\cref{eq: first_order_D0D4}, and then using the Clifford algebra and the first-order equation of motion \cref{eq: first_order_D0D4}, it is possible 
to re-express some of the terms appropriately such that we derive an ordinary, second-order differential equation of the following form
\eq$\label{eq: D0D4_sec_ord_eq}
\left( \mathcal{A}^{\pm}_{1} \partial_{\varrho}^2 + \mathcal{A}^{\pm}_{2} \partial_{\varrho} + \mathcal{A}^{\pm}_{3} \bar{M}^2 +\mathcal{A}^{\pm}_{4}
 \Gamma^{\varrho} + \mathcal{A}^{\pm}_{5} \right) \psi^{\ell^{\pm}}(\varrho) = 0\,,$ 
where we have used the decomposition $\Psi^{\ell^{\pm}}(x^\lambda,\rho)=e^{i\,k_\lambda x^\lambda}\,\psi^{\ell^{\pm}}(\rho)$ and the relation $M^2=-\eta^{\mu\nu}k_\mu k_\nu$.\footnote{This particular
decomposition results to
\gatt$\Gamma^\mu\Gamma^\nu\pa_\mu\pa_\nu \Psi^{\ell^{\pm}}=-k_\mu k_\nu \Gamma^\mu\Gamma^\nu\Psi^{\ell^{\pm}}=-k_\mu k_\nu
\left(2\eta^{\mu\nu}\mathbb{I}_{32}-\Gamma^\nu\Gamma^\mu\right)\Psi^{\ell^{\pm}}=2M^2\,\Psi^{\ell^{\pm}}+k_\mu k_\nu\Gamma^\nu\Gamma^\mu
\Psi^{\ell^{\pm}}\xRightarrow{\mu\leftrightarrow\nu}\nonum\\[1mm]
-k_\mu k_\nu \Gamma^\mu\Gamma^\nu\Psi^{\ell^{\pm}}=2M^2\,\Psi^{\ell^{\pm}}+k_\nu k_\mu \Gamma^\mu\Gamma^\nu\Psi^{\ell^{\pm}}\Ra
-k_\mu k_\nu \Gamma^\mu\Gamma^\nu\Psi^{\ell^{\pm}}=M^2\,\Psi^{\ell^{\pm}}\Ra\nonum\\[1mm]
\Gamma^\mu\Gamma^\nu\pa_\mu\pa_\nu \Psi^{\ell^{\pm}}=M^2\,\Psi^{\ell^{\pm}}\,.$} In addition, we have also used the dimensionless quantities $\varrho = \rho/L$, $\bar{M}^2 =M^2 R^7/L^5$. The $\mathcal{A}$ factors which appropriately describe the second-order equation are shown below:
\begin{equation} \label{eq: D0D4factors}
\begin{split}
\mathcal{A}^{\pm}_1 &= 1, \quad \mathcal{A}^{\pm}_2 = \frac{3}{\varrho} + \frac{21}{4} \frac{\varrho}{\varrho^2 +1}, \quad \mathcal{A}^{\pm}_3 = 
\frac{1}{\left(\varrho^2 +1 \right)^{7/2}}\,, \\
\mathcal{A}^{\pm}_4 &= \frac{7}{4} \frac{1\pm 2 \ell \pm 3}{\varrho^2+1} \mp \frac{1}{\varrho^2} \left( \ell + \frac{3}{2} \right) + \frac{21}{8} 
\frac{\varrho^2}{\left( \varrho^2 + 1 \right)^2}\,, \\
\mathcal{A}^{\pm}_5 &= \frac{7}{4} \frac{1}{\varrho^2+1} \left( 5 \mp 3 \mp 2 \ell \right) - \frac{63}{64} \frac{\varrho^2}{(\varrho^2+1)^2} - 
\frac{1}{\varrho^2} \left( \ell^2 + 3 \ell+\frac{3}{2} \right)\,.
\end{split}
\end{equation}
In the above equations, the plus sign in the $\mathcal{A}$ factors denotes the positive eigenvalue of the spinorial harmonics ($\mathcal{G}$ operators) 
and the negative sign superscript is related to the negative eigenvalue ($\mathcal{F}$ operators). Moreover, the projection $\Gamma^\rho \Psi=\pm\Psi$
(or $\Gamma^\varrho \psi=\pm\psi$) generates two additional second-order differential equations when it is applied to \cref{eq: D0D4_sec_ord_eq}. 
The spinor which under the aforementioned projection remains invariant will be denoted by $\Psi^{\oplus}$ (or $\psi^\oplus$), while $\Psi^{\ominus}$ 
(or $\psi^\ominus$) will represent the spinor which under the same projection goes to minus itself. Therefore, we can distinguish the following cases:
\begin{enumerate}[(i)]
\item Positive spinorial harmonics eigenvalues:
\begin{itemize}
\item Positive $\Gamma^\varrho$-projection:
\eq$\left( \mathcal{A}^{+}_{1} \partial_{\varrho}^2 + \mathcal{A}^{+}_{2} \partial_{\varrho} +\mathcal{A}^{+}_{3} \bar{M}^2 +\mathcal{A}^{+}_{4} + 
\mathcal{A}^{+}_{5}\right) \psi^{\oplus}_{\calG_\ell}(\varrho) = 0\Ra\nonum$
\bal$\label{eq: pos_sph_pos_proj_04}
\left[\pa^2_\varrho+\left(\frac{3}{\varrho}+\frac{21}{4}\frac{\varrho}{\varrho^2+1}\right)\pa_\varrho+\frac{\bar{M}^2}{(\varrho^2+1)^{7/2}}\right.&-
\frac{1}{\varrho^2}(\ell^2+4\ell+3)\nonum\\
&\left.+\frac{21}{2}\frac{1}{\varrho^2+1}+\frac{105}{64}\frac{\varrho^2}{(\varrho^2+1)^2}\right]\psi^{\oplus}_{\calG_\ell}(\varrho) = 0\,.$
\item Negative $\Gamma^\varrho$-projection:
\eq$\left( \mathcal{A}^{+}_{1} \partial_{\varrho}^2 + \mathcal{A}^{+}_{2} \partial_{\varrho} + \mathcal{A}^{+}_{3} \bar{M}^2 -\mathcal{A}^{+}_{4} + 
\mathcal{A}^{+}_{5}\right) \psi^{\ominus}_{\calG_\ell}(\varrho) = 0\Ra\nonum$
\bal$\label{eq: pos_sph_neg_proj_04}
\left[\pa^2_\varrho+\left(\frac{3}{\varrho}+\frac{21}{4}\frac{\varrho}{\varrho^2+1}\right)\pa_\varrho+\frac{\bar{M}^2}{(\varrho^2+1)^{7/2}}\right.&-
\frac{1}{\varrho^2}(\ell^2+2\ell)\nonum\\
&\left.-\frac{7}{2}\frac{1+2\ell}{\varrho^2+1}-\frac{231}{64}\frac{\varrho^2}{(\varrho^2+1)^2}\right]\psi^{\ominus}_{\calG_\ell}(\varrho) = 0\,.$
\end{itemize} 
\item Negative spinorial harmonics eigenvalues:
\begin{itemize}
\item Positive $\Gamma^\varrho$-projection:
\eq$\left( \mathcal{A}^{-}_{1} \partial_{\varrho}^2 + \mathcal{A}^{-}_{2} \partial_{\varrho} + \mathcal{A}^{-}_{3} \bar{M}^2 +\mathcal{A}^{-}_{4} + 
\mathcal{A}^{-}_{5}\right) \psi^{\oplus}_{\calF_\ell}(\varrho) = 0\Ra\nonum$
\bal$\label{eq: neg_sph_pos_proj_04}
\left[\pa^2_\varrho+\left(\frac{3}{\varrho}+\frac{21}{4}\frac{\varrho}{\varrho^2+1}\right)\pa_\varrho+\frac{\bar{M}^2}{(\varrho^2+1)^{7/2}}\right.&-
\frac{1}{\varrho^2}(\ell^2+2\ell)\nonum\\
&\left.+\frac{21}{2}\frac{1}{\varrho^2+1}+\frac{105}{64}\frac{\varrho^2}{(\varrho^2+1)^2}\right]\psi^{\oplus}_{\calF_\ell}(\varrho) = 0\,.$
\item Negative $\Gamma^\varrho$-projection:
\eq$\left( \mathcal{A}^{-}_{1} \partial_{\varrho}^2 + \mathcal{A}^{-}_{2} \partial_{\varrho} + \mathcal{A}^{-}_{3} \bar{M}^2 -\mathcal{A}^{-}_{4} + 
\mathcal{A}^{-}_{5}\right) \psi^{\ominus}_{\calF_\ell}(\varrho) = 0\Ra\nonum$
\bal$\label{eq: neg_sph_neg_proj_04}
\left[\pa^2_\varrho+\left(\frac{3}{\varrho}+\frac{21}{4}\frac{\varrho}{\varrho^2+1}\right)\pa_\varrho+\frac{\bar{M}^2}{(\varrho^2+1)^{7/2}}\right.&-
\frac{1}{\varrho^2}(\ell^2+4\ell+3)\nonum\\
&\left.+\frac{7}{2}\frac{5+2\ell}{\varrho^2+1}-\frac{231}{64}\frac{\varrho^2}{(\varrho^2+1)^2}\right]\psi^{\ominus}_{\calF_\ell}(\varrho) = 0\,.$
\end{itemize}
\end{enumerate}

\begin{figure}[t]
	\begin{center}
	\begin{subfigure}[t]{0.44\textwidth}
	\includegraphics[width=\textwidth]{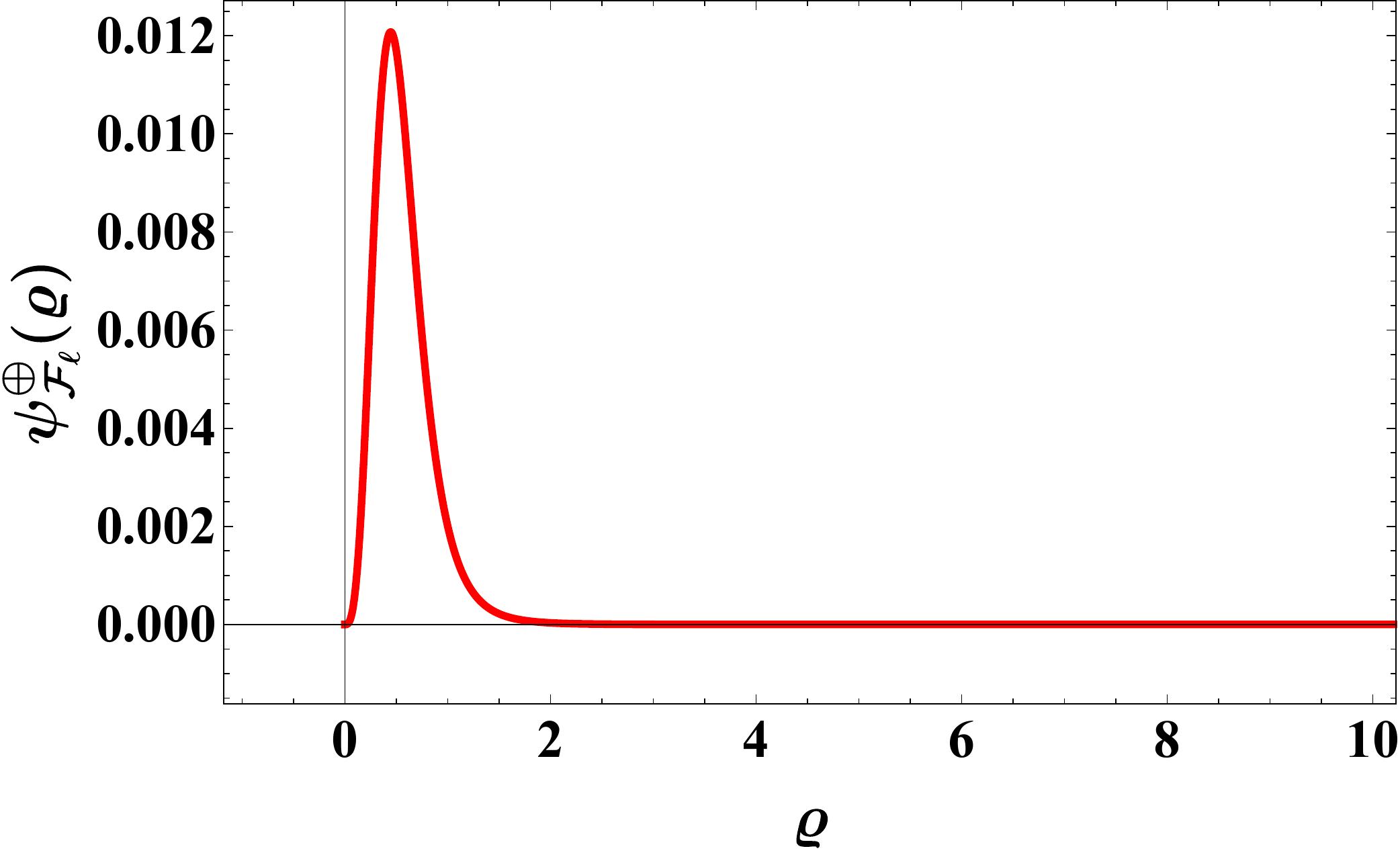}
	\caption{\hspace*{-4em}}
	\label{fig: plotFposIR}
	\end{subfigure}
	\hspace{1em}
	\begin{subfigure}[t]{0.478\textwidth}
	\includegraphics[width=\textwidth]{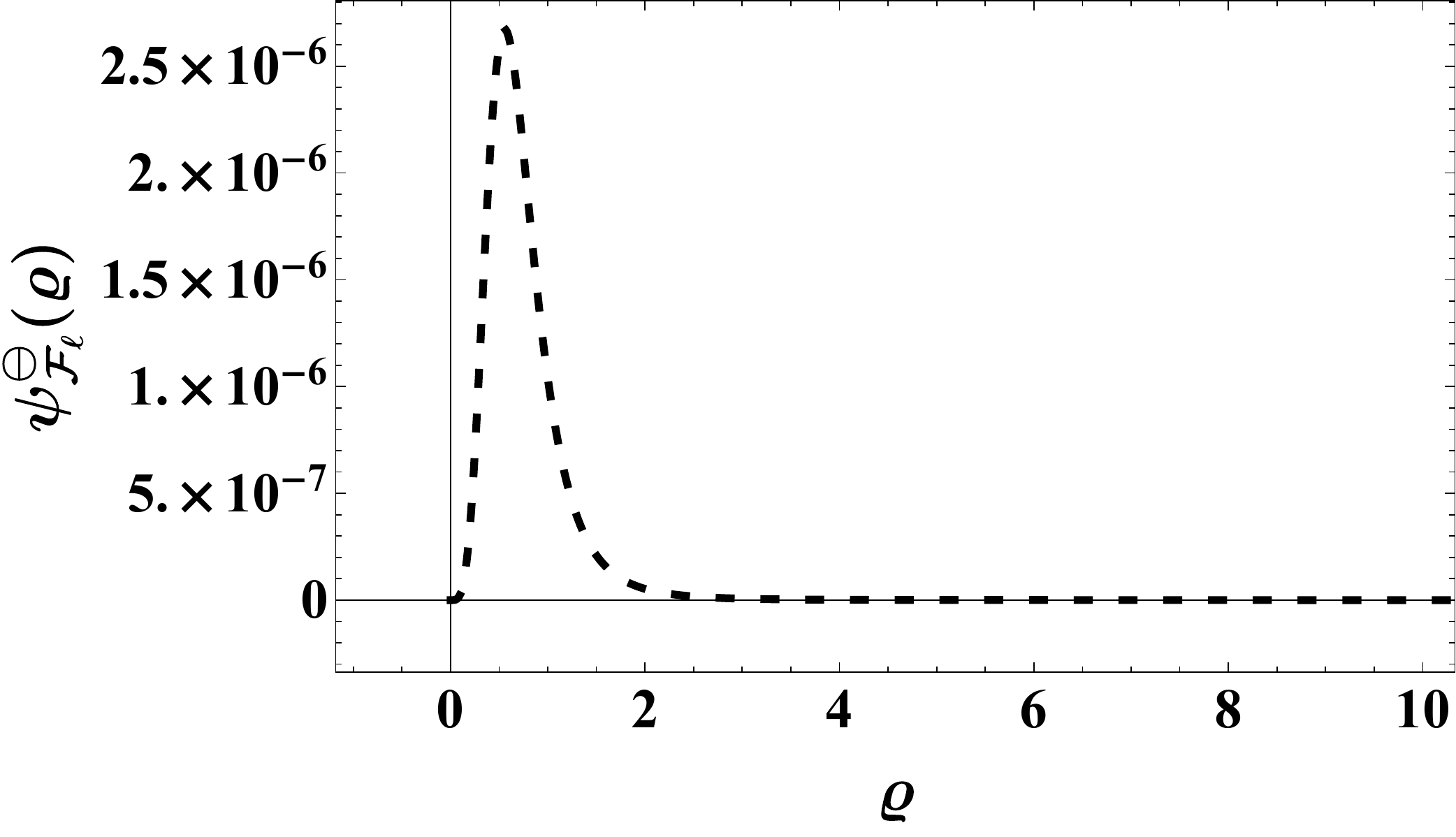}
	\caption{\hspace*{-6em}}
	\label{fig: plotFnegIR}
	\end{subfigure}
	\vspace{-0.5em}
	\caption{(a) The solution obtained from the positive eigenvalue of the $\Gamma^{\varrho}$ projector, while (b) refers to the negative eigenvalue of the
	$\Gamma^{\varrho}$ projector. Both of them are dual to the field theory $\mathcal{F}$ modes for quantum numbers $n=0, \ell=3$.}
	\vspace{-1em}
	\label{plotFIR}
  	\end{center}
\end{figure}

\begin{figure}[h]
	\begin{center}
	\begin{subfigure}[t]{0.48\textwidth}
	\includegraphics[width=\textwidth]{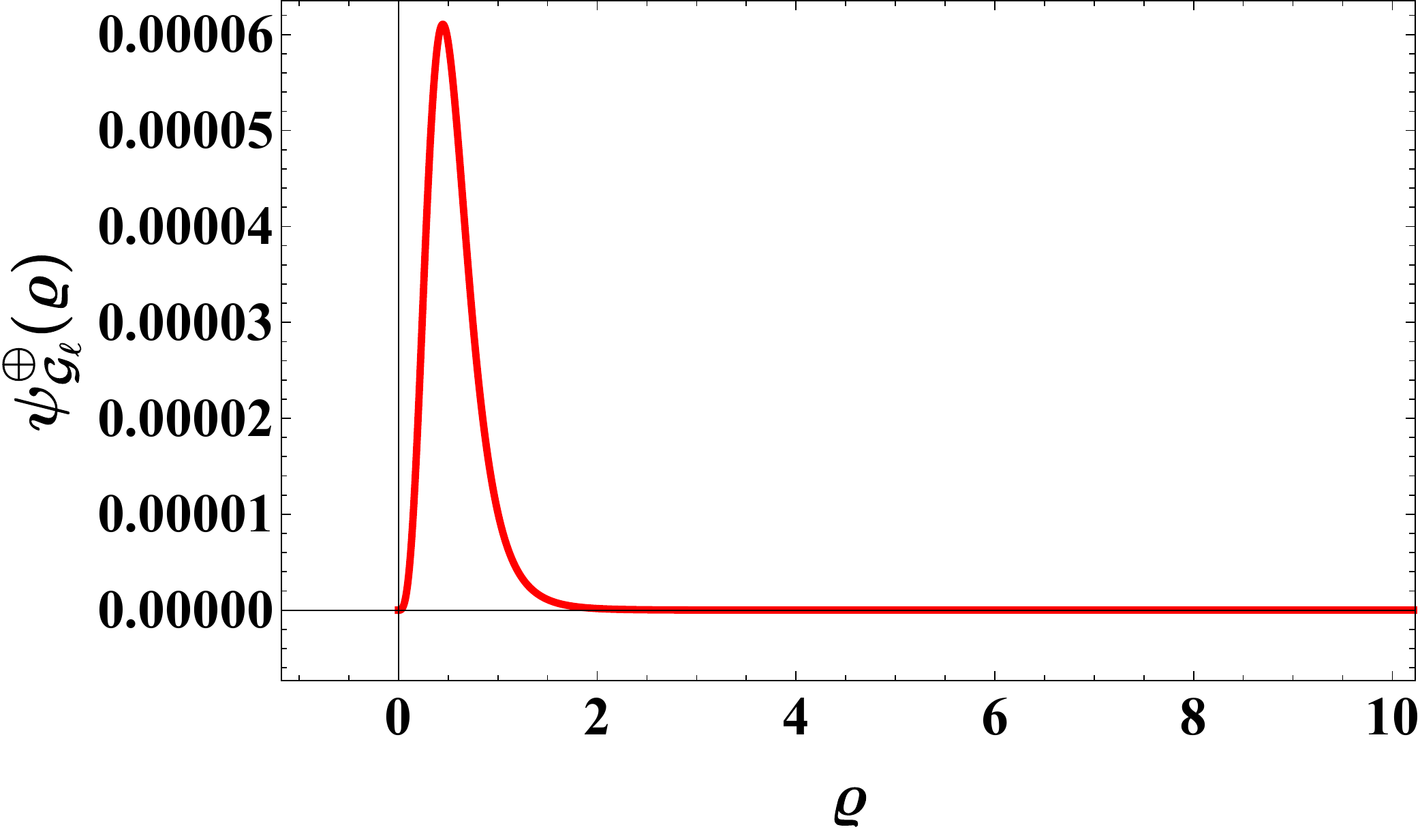}
	\caption{\hspace*{-5.1em}}
	\label{fig: plotGposIR}
	\end{subfigure}
	\hspace{1em}
	\begin{subfigure}[t]{0.45\textwidth}
	\includegraphics[width=\textwidth]{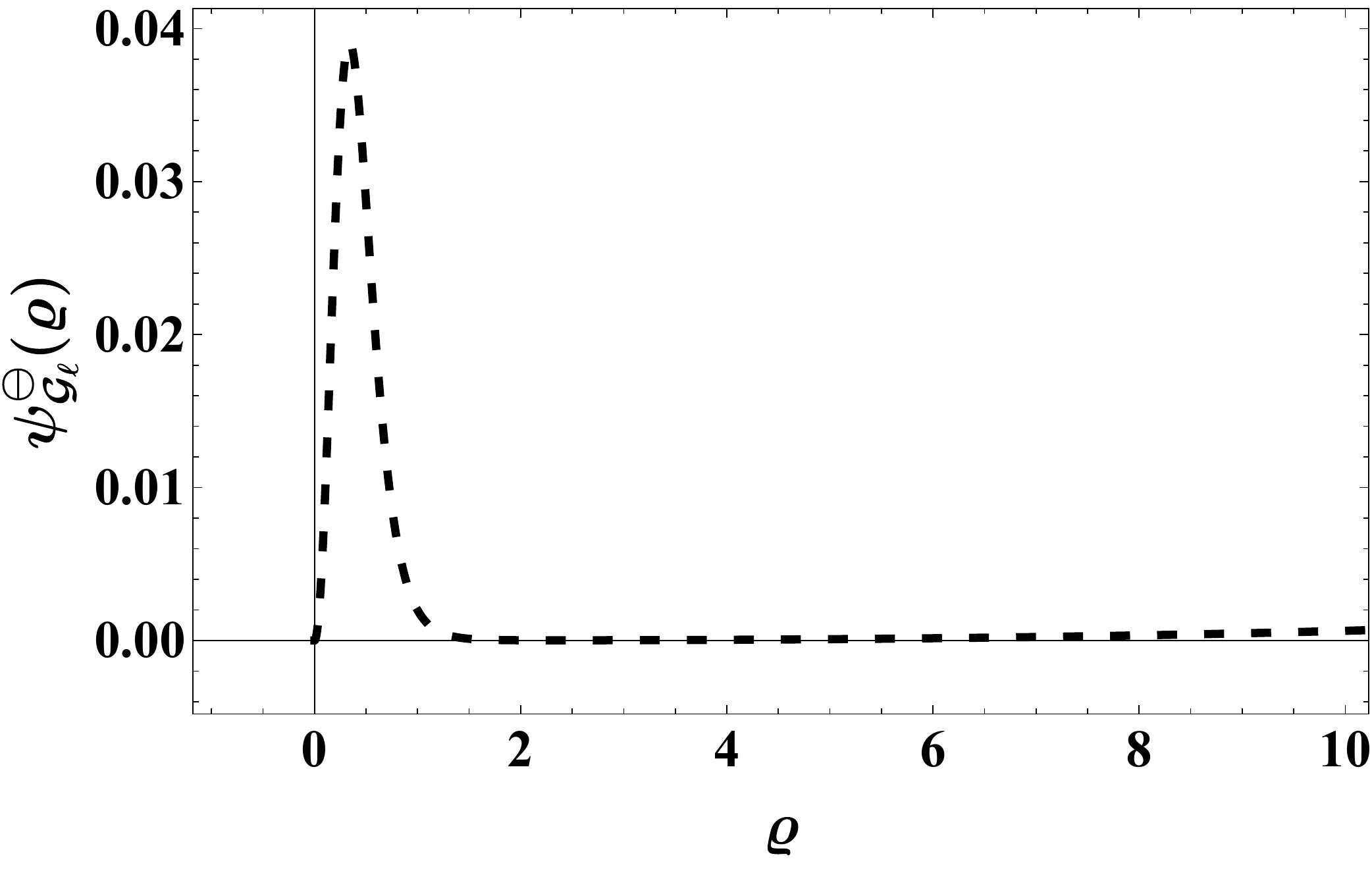}
	\caption{\hspace*{-3.5em}}
	\label{fig: plotGnegIR}
	\end{subfigure}
	\vspace{-0.5em}
	\caption{(a) The solution obtained from the positive eigenvalue of the $\Gamma^{\varrho}$ projector, while (b) refers to the negative eigenvalue of the
	$\Gamma^{\varrho}$ projector. Both of them are dual to the field theory $\mathcal{G}$ modes for quantum numbers $n=0, \ell=2$.}
	\vspace{-1em}
	\label{plotGIR}
  	\end{center}
\end{figure}

It is straightforward to see that applying $p=0$ in \cref{eq: bosonic_spectra_4} yields the corresponding bosonic equation of motion for the 
D$0$/D$4$-brane system, namely, it is
\eq$\label{eq: bos_eq_04}
\partial^2_{\varrho}\, f_\ell(\varrho) + \frac{3}{\varrho}\, \partial_{\varrho}f_\ell(\varrho) + \left[ \frac{\bar{M}^2}{(1 + \varrho^2)^{7/2}} - 
\frac{\ell(\ell+2)}{\varrho^2} \right]f_\ell(\varrho) = 0\,.$
One can easily verify that the transformation
\eq$\label{eq: pos_sph_trans_04}
\psi^{\oplus}_{\calG_{\ell}}(\varrho)=\frac{1}{(1+\varrho^2)^{21/16}}\, f_{\ell+1}(\varrho)\,,$ 
maps \cref{eq: pos_sph_pos_proj_04} to \cref{eq: bos_eq_04} for $\ell \ra\ell+1$, while the transformation
\eq$\label{eq: neg_sph_trans_04}
\psi^{\oplus}_{\calF_{\ell}}(\varrho)=\frac{1}{(1+\varrho^2)^{21/16}}\, f_\ell(\varrho)\,,$
maps \cref{eq: neg_sph_pos_proj_04} to \cref{eq: bos_eq_04}. Notice that the above transformations can be obtained from \eqref{eq: basic_1} 
by setting $p=0$. In contrast, eqs. \eqref{eq: pos_sph_neg_proj_04} and \eqref{eq: neg_sph_neg_proj_04} cannot be mapped to \cref{eq: bos_eq_04}. 
This particular behaviour is anticipated due to the structural difference of the two separate equations of motion obtained by the application of the two 
different projection eigenvalues of the $\Gamma^\rho$ matrix. In order to demonstrate their difference, we simply plot the mode-solutions from which 
we obtain the correct mass squared eigenvalues, see Figures \ref{plotFIR} and \ref{plotGIR}. At this point, it is important to mention that the
quantum number $n$ counts the nodes of the functions $\psi^{\oplus, \ominus}_{\calG_\ell}$ and $\psi^{\oplus, \ominus}_{\calF_\ell}$; it is clear that
in both Figures \ref{plotFIR} and \ref{plotGIR} the quantum number $n$ is zero.
From these plots it is also very clear how 
we must shift appropriately the quantum number $\ell$ such that all the states are part of the same massive supermultiplet as dictated in 
\cite{Kruczenski:2003be}. We have employed this shift in the $\ell$ quantum number in all the tables with our numerical results such that we keep our 
presentation short.
While we have provided appropriate transformations such that we analytically map the fermionic degrees of freedom to the bosonic ones, 
\cref{eq: pos_sph_trans_04,eq: neg_sph_trans_04}, here we demonstrate the numerical approach that we used to determine the mass eigenvalues 
directly from the second order equations of motion for the D$0$/D$4$ brane junction. The same numerical approach was of course used to compute 
the solutions in Figures \ref{plotFIR} and \ref{plotGIR}.

\begin{itemize}
\item We are shooting from the $\Lambda_{\rm{IR}}$ to the $\Lambda_{\rm{UV}}$ by fine-tuning $\bar{M}^2$ such that the mode solutions are 
normalizable in the UV and small in amplitude \cite{Kruczenski:2003be}. 
\item We use  $\Lambda_{\rm{IR}} = 10^{-7}$ and  $\Lambda_{\rm{UV}} = 10$.
\item For the initial conditions we use 
\begin{equation}
\begin{split}
\psi_{\mathcal{G},+} (\varrho)|_{\varrho \rightarrow 0} &= \varrho^{\ell+1}, \hspace{1cm} \partial_{\varrho} \psi_{\mathcal{G},+} (\varrho)|_{\varrho 
\rightarrow 0} = (\ell+1) \varrho^{\ell}\,, \\
\psi_{\mathcal{G},-} (\varrho)|_{\varrho \rightarrow 0} &= \varrho^{\ell}, \hspace{1.5cm} \partial_{\varrho} \psi_{\mathcal{G},-} (\varrho)|_{\varrho 
\rightarrow 0} = \ell \varrho^{\ell-1}\,,
\end{split}
\end{equation} 
 
\begin{equation}
\begin{split}
\psi_{\mathcal{F},+} (\varrho)|_{\varrho \rightarrow 0} &= \varrho^{\ell}, \hspace{1.5cm} \partial_{\varrho} \psi_{\mathcal{F},+} (\varrho)|_{\varrho 
\rightarrow 0} = \ell \varrho^{\ell-1}\,, \\
\psi_{\mathcal{F},-} (\varrho)|_{\varrho \rightarrow 0} &= \varrho^{\ell+1}, \hspace{1cm} \partial_{\varrho} \psi_{\mathcal{F},-} (\varrho)|_{\varrho 
\rightarrow 0} = (\ell+1) \varrho^{\ell}\,.
\end{split}
\end{equation} 
\end{itemize}
We show explicitly the first few numerical values of the masses in \Cref{tab: d0}. The numerical values for $n$ and $\ell$ describe the $\mathcal{F}$ 
operators and one needs to shift appropriately the $\ell$ to read off the value of the $\mathcal{G}$ states.
\begin{table}[H]
    \centering
   \begin{tabular}{|c|c|c|c|c|c|}
   \cline{3-6}
    \multicolumn{2}{c}{}     & \multicolumn{4}{|c|}{$\boldsymbol{n}$} \\ \cline{3-6}
    \multicolumn{2}{c|}{}        & 0 & 1 & 2 & 3       \\ \hline
    \multirow{4}{0.5em}{$\boldsymbol{\ell}$}  
    						  & 0 & 18.164  & 62.390 & 132.57 & 228.71 \\ \cline{2-6}  
                             & 1 & 52.188 & 115.60 & 204.53 & 319.13 \\ \cline{2-6}
                             & 2 & 102.53 & 185.35 & 293.52 & 422.96 \\ \cline{2-6}
                             & 3 & 168.85 & 271.32 & 398.78 & 551.65 \\ \cline{1-6}
  \end{tabular}
\caption{Numerical results of the parameter $\bar{M}^2$ for the D$0$/D$4$ brane junction.}
\label{tab: d0}
\end{table}
We have also explicitly checked that we obtain the same spectrum from 
\cref{eq: bos_eq_04}. In order to perform the numerical analysis for the bosonic degrees of freedom we used as initial conditions $f(\varrho)
|_{\varrho \rightarrow 0} = \varrho^{\ell}$ and $\partial_{\varrho} f(\varrho)|_{\varrho \rightarrow 0} = \ell \varrho^{\ell-1}$ 
\cite{Myers:2006qr, Arean:2006pk} and the other parameters are the same as previously.

\begin{figure}[H]
	\begin{center}
	\begin{subfigure}[t]{0.47\textwidth}
	\includegraphics[width=\textwidth]{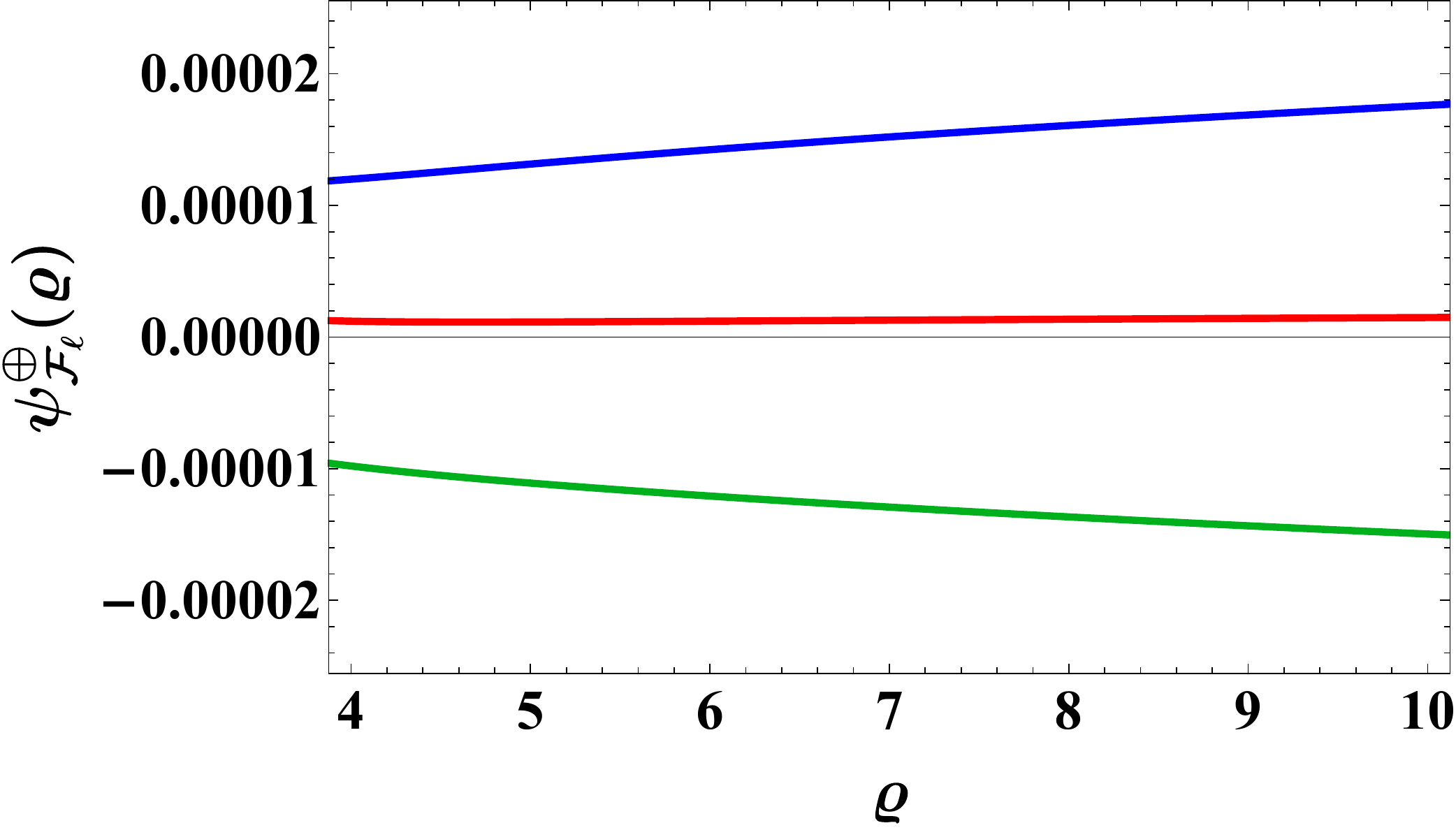}
	\caption{\hspace*{-5.6em}}
	\label{fig: plotFposUV}
	\end{subfigure}
	\hspace{1em}
	\begin{subfigure}[t]{0.47\textwidth}
	\includegraphics[width=\textwidth]{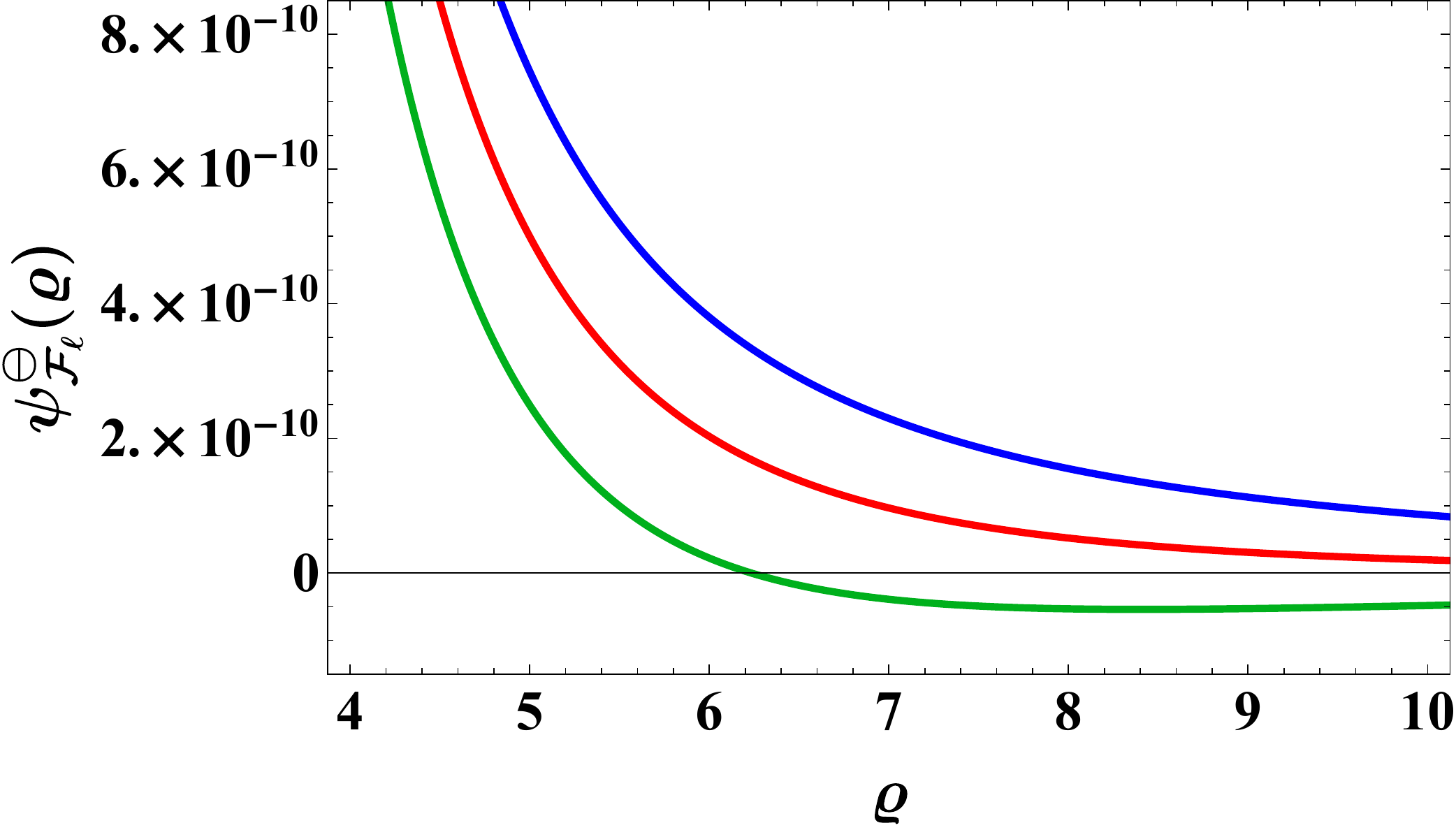}
	\caption{\hspace*{-5.6em}}
	\label{fig: plotFnegUV}
	\end{subfigure}
	\vspace{-0.5em}
	\caption{For both (a) and (b), the masses are $\bar{M}^2=169$ (blue), $169.0736$ (red), $169.1$ (green) (from top to bottom). 
	The parameters $n$ and $\ell$ are the same as in Figure \ref{plotFIR}, namely $n=0$, $\ell=3$. We have zoomed 
	in to the plots in order to show the behaviour near the UV cut-off. Clearly the red line, which is the correct solution, can be uniquely determined from 
	either  positive or negative $\Gamma^{\varrho}$-projections.}
	\vspace{-1em}
	\label{plotFUV}
  	\end{center}
\end{figure}

\begin{figure}[H]
	\begin{center}
	\begin{subfigure}[t]{0.49\textwidth}
	\includegraphics[width=\textwidth]{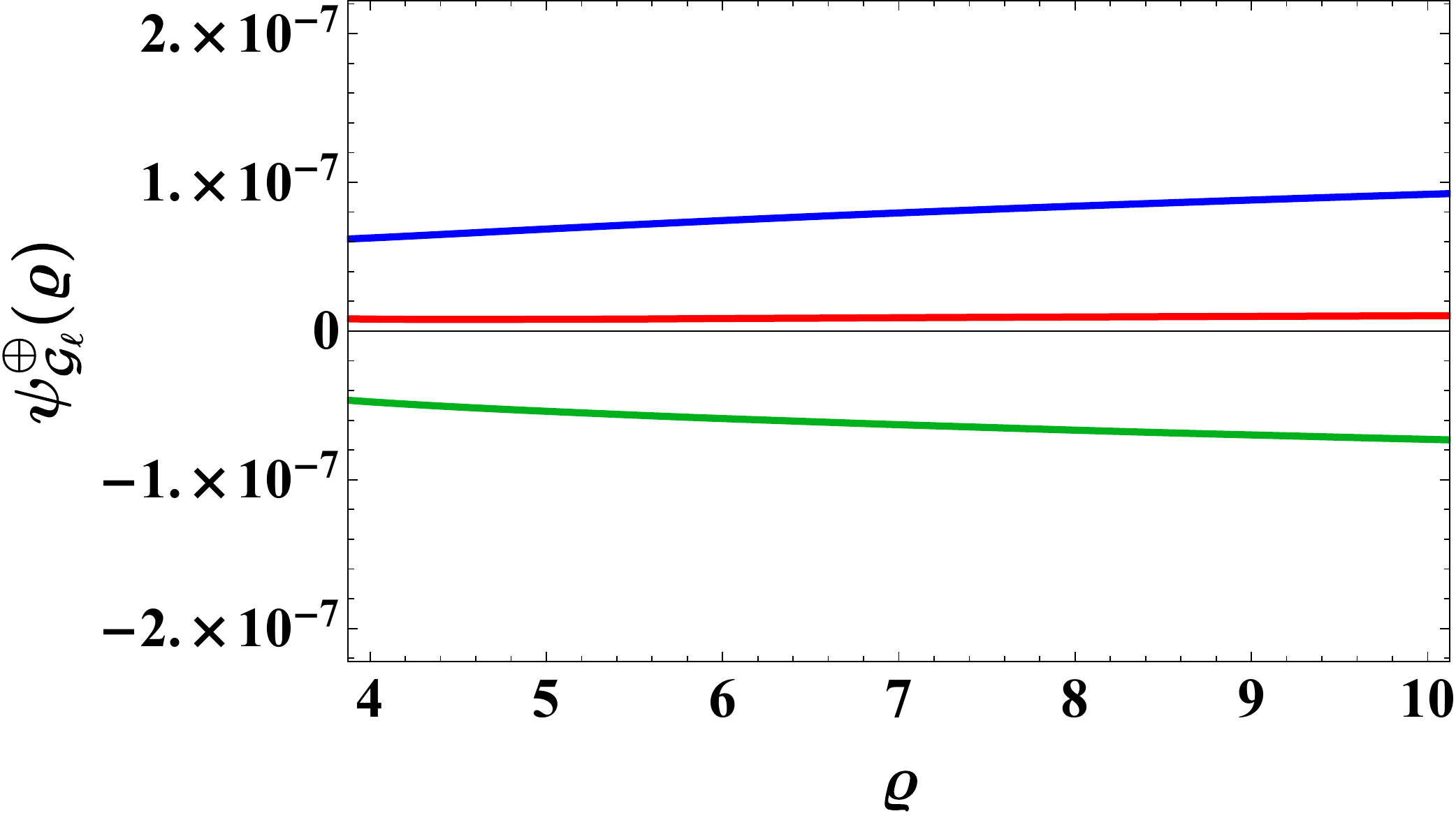}
	\caption{\hspace*{-5.95em}}
	\label{fig: plotGposUV}
	\end{subfigure}
	\hspace{1em}
	\begin{subfigure}[t]{0.47\textwidth}
	\includegraphics[width=\textwidth]{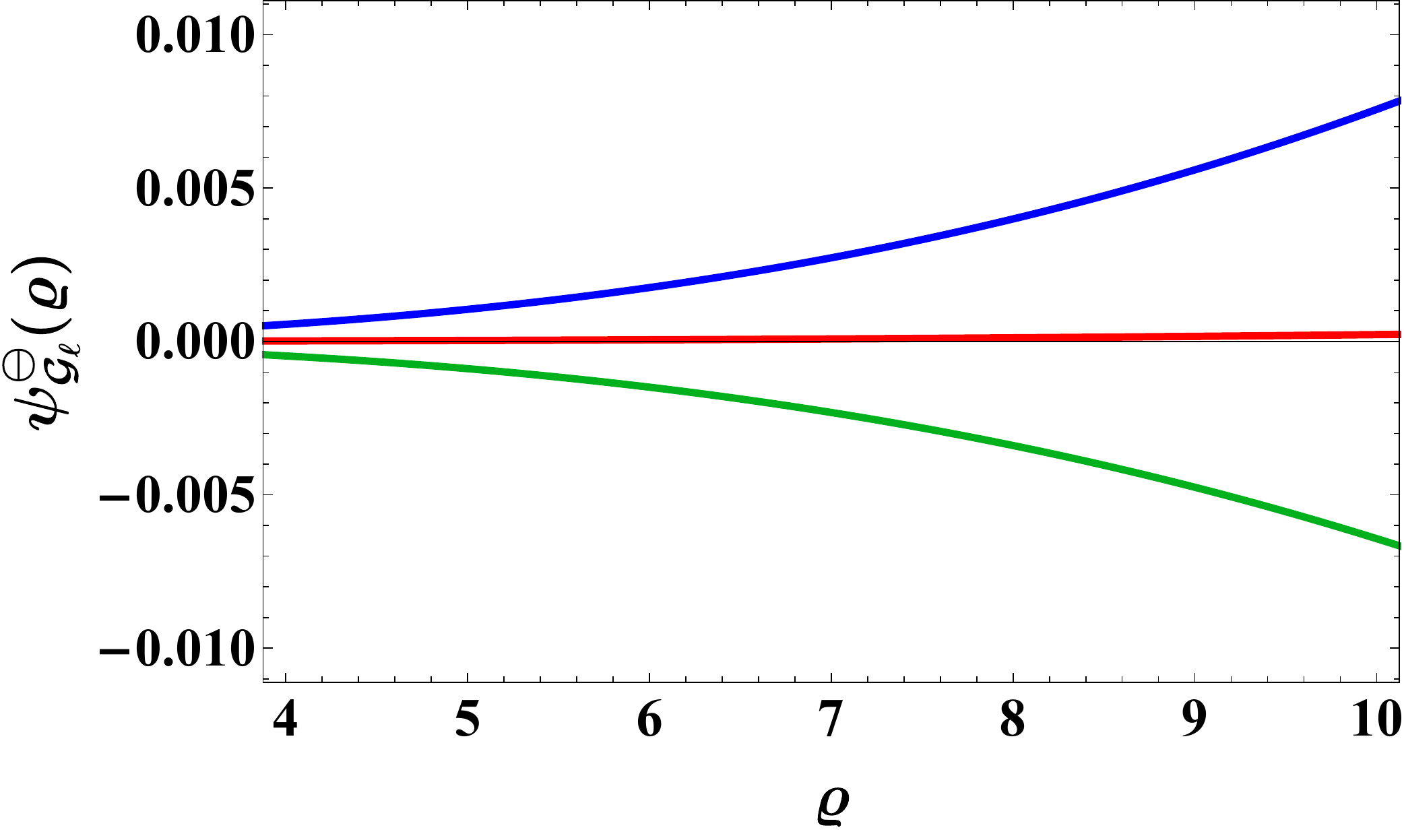}
	\caption{\hspace*{-4.7em}}
	\label{fig: plotGnegUV}
	\end{subfigure}
	\vspace{-0.5em}
	\caption{For both (a) and (b), the masses are $\bar{M}^2=169$ (blue), $169.0736$ (red), $169.1$ (green) (from top to bottom). The parameters
	$n$ and $\ell$ are the same as in Figure \ref{plotGIR}, namely $n=0$, $\ell=2$. We have zoomed in to the plots in order to show the behaviour near the 
	UV cut-off. Clearly the red line, which is the correct solution, can be uniquely determined from either 
	positive or negative $\Gamma^{\varrho}$-projections.}
	\vspace{-1em}
	\label{plotGUV}
  	\end{center}
\end{figure}

As a final project within the D$0$/D$4$ brane setup, we are demonstrating explicitly that the $\bar{M}^2$ value can be obtained from either 
the positive or the negative eigenvalues of the $\Gamma^{\varrho}$ matrix. This statement holds true of course for both the string modes dual to 
the $\mathcal{F}$ and $\mathcal{G}$ modes. 
In order to do so, we predict the mass of the $n=0,\ \ell=2$ for the supergravity modes dual to the $\mathcal{G}$ operators (the ones associated with the 
positive eigenvalue of the spinor spherical harmonics) and the $n=0,\ \ell=3$ state of the fluctuation dual to the $\mathcal{F}$ operators (related to the 
negative eigenvalue of the spinor harmonics). Note that these particular  $\mathcal{F}$ and  $\mathcal{G}$ states should have the same mass, since we 
are shifting $\ell$ by one. We present our results for the $\mathcal{F}$ modes in \Cref{plotFUV} and we proceed to the results related to the 
$\mathcal{G}$ modes, which are shown in \Cref{plotGUV}.
\section{The general procedure} \label{sec: general idea}
In this section, we present the necessary mathematical machinery for computing the equation of motion and solve for the mass 
eigenvalues of the world-volume fermions in all possible probe-brane setups. The main task at hand, in order to determine the equations
of motion and solve for the mass eigenvalues, is to derive a second-order ordinary differential equation of a scalar function depending on the holographic 
radial coordinate. To do so, we: 
\begin{itemize}
\item Compute the vielbein and the spin-connection components of the background geometry from eqs. \eqref{eq: vielbeine-1}-\eqref{eq: vielbeine-3}
and \eqref{eq: spinconnection-1}-\eqref{eq: spinconnection-4}.

\item Compute the dilaton field and the R-R field strength from \cref{eq: dil&RRpot}. In addition, we re-express the R-R field strength components
in the vielbein basis.

\item Write the curved $\Gamma$-matrices in terms of flat ones. It is also worthwhile commenting that we do not need an explicit representation of these 
matrices as we will be dealing only with projectors. These can be computed by means of the Clifford algebra and general properties that the generators of 
the algebra satisfy.

\item Obtain the equation of motion for the ten-dimensional spinor $\Psi$ by varying the fermionic DBI action with respect to the conjugate spinor 
$\bar{\Psi}$.
\item Determine the first-order differential equation of the spinor using the explicit expression of the Dirac operator given by \cref{eq: Dkdirac}
or \eqref{eq: Dkdirac-sec}.
For the spherical piece we can utilize the spinor spherical harmonics \cite{Camporesi:1995fb}
\begin{equation} \label{eq: spinor harmonics}
\slashed{D}_{S^{k-d-1}}\Psi^{\ell^{\pm}} = \pm \left(\ell + \frac{k-d-1}{2} \right)\Psi^{\ell^{\pm}}\,,
\end{equation}
while, for the bulk piece, we use separation of variables for the Minkowski coordinates and the radial coordinate by making a plane-wave ansatz, 
with the wave-vector satisfying $k^\mu k_\mu=-M^2$ and $M$ being the mass of the fluctuation.
 
It is also important to stress here that performing the calculations in the vielbein basis $\{e^{(M)}\}$, instead of the usual $\{dx^M\}$, will affect all 
spacetime derivatives in the spinor equation of motion. Thus, in the D$p$/D$k$-brane systems in which the dilaton field $\phi$ does not vanish, we should 
keep in mind that $\Gamma^M\pa_M\phi$ in the $\{dx^M\}$ basis will become $\Gamma^N e_{(N)}{}^M\pa_{M}\phi$ in the $\{e^{(M)}\}$ basis. From 
\cref{eq: dil&RRpot} and after a straightforward calculation, we find that
\eq$\label{eq: dilterm}
\Gamma^N e_{(N)}{}^M\pa_{M}\phi=-\frac{(7-p)(3-p)}{4}\frac{1}{R^{\frac{7-p}{4}}}\frac{1}{r^{\frac{p+1}{4}}}\left(\rho\,\Gamma^\rho+
\sum_{m=1}^{9-p-k+d}w^m\,\Gamma_{\wtilde{m}}\right)\,.$
Subtracting now \cref{eq: dilterm} from \eqref{eq: Dkdirac-sec} we get
\bal$\label{eq: dirac&dilterm}
\slashed{D}_{Dk}\Psi^{\ell^{\pm}}-\frac{\Gamma^N e_{(N)}{}^M\pa_{M}\phi}{2}\,\Psi^{\ell^{\pm}}=&\left\{\frac{R^{\frac{7-p}{4}}}{r^{\frac{7-p}{4}}}\,
\Gamma^{\mu}  \partial_{\mu} +\frac{r^{\frac{7-p}{4}}}{R^{\frac{7-p}{4}}}\, \Gamma^{\rho}  \partial_{\rho} \pm\frac{1}{\rho}
\frac{r^{\frac{7-p}{4}}}{R^{\frac{7-p}{4}}} \left(\ell + \frac{m}{2} \right)\right.\nonum\\[2mm]
&\left.+\frac{1}{R^{\frac{7-p}{4}}}\left[ \frac{7-p}{8} \frac{\rho}{r^{\frac{p+1}{4}}}+\frac{m}{2}\frac{r^{\frac{7-p}{4}}}{\rho}\right]  \Gamma^{\rho}
\right\}\Psi^{\ell^{\pm}}\,,$
where we have defined $m\equiv k-d-1$.

\item Once we obtain the first-order equation of motion that the spinor satisfies we act with the differential operator 
\begin{equation}\label{eq: oper_to_secDE}
\frac{R^{\frac{7-p}{4}}}{r^{\frac{7-p}{4}}}\,\Gamma^{\mu}  \partial_{\mu} + \frac{r^{\frac{7-p}{4}}}{R^{\frac{7-p}{4}}}\, \Gamma^{\rho}  
\partial_{\rho}\,,
\end{equation}
on this first-order equation of motion.
Then, by using some basic identities of the Clifford algebra and after some tedious algebra we derive a second-order ordinary differential equation of the following 
schematic form 
\begin{align} \label{eq: generalednsetup}
\left( \mathcal{A}_{1}\, \partial_{\varrho}^2 + \mathcal{A}_{2}\, \partial_{\varrho} + \mathcal{A}_{3}\, \bar{M}^2 +\mathcal{A}_{4}\, \Gamma^{
\varrho} + \mathcal{A}_{5} \right) \psi(\varrho) = 0\,,
\end{align} 
where we have also used the  dimensionless quantities 
\begin{align}\label{eq: gen_dim-less_quants}
\varrho &= \frac{\rho}{L}\,, &&& \bar{M}^2 =\frac{M^2 R^{7-p}}{L^{5-p}}\,.
\end{align}
In the above, $L^2=\sum_{m=1}^{9-p-k+d}(w^m)^2$ constitutes the distance between the D$p$- and the D$k$-branes. $\mathcal{A}$'s are the 
coefficients of interest that we compute explicitly for each case we consider.
\end{itemize}
\section{Fermions in the D1-background} \label{sec: D1}
The branes in the D$1$-background are arranged in the following way, see \Cref{table: D1_brane_intersections}. 
\begin{table*}[h]
\begin{center}
\begin{tabular}{ |c|c|c|c|c|c|c|c|c|c|c|c|}
 \hline
 &&&&&&&&&&\\[-0.95em] 
   									& $x^0$ & $x^1$ & $x^2$ & $x^3$ & $x^4$ & $x^5$ & $x^6$ & $x^7$ & $x^8$ & $x^9$			\\ 
 \hline
 Background D$1$-brane 	& --- & --- & $\bullet$ & $\bullet$ & $\bullet$ & $\bullet$ & $\bullet$ & $\bullet$ & $\bullet$ & $\bullet$	\\ 
 \hline
 Probe D$3$-brane 			& --- & $\bullet$ & --- & --- & --- & $\bullet$ & $\bullet$ & $\bullet$ & $\bullet$ & $\bullet$				\\
 \hline
 Probe D$5$-brane 			& --- & --- & --- & --- & --- & --- & $\bullet$ & $\bullet$ & $\bullet$ & $\bullet$				\\
 \hline
\end{tabular}
\caption{The supersymmetric brane intersections. In the above notation --- denotes that a brane extends along that particular direction, 
while $\bullet$ means that the coordinate is transverse to the brane.}
\label{table: D1_brane_intersections}
\end{center}
\end{table*}

The fermionic action is given by \footnote{We have written the action using single spinor notation rather than the two-component one. In order to do so
we have manipulated appropriately some Pauli matrices as explained in Appendix of paper \cite{Martucci:2005rb}.}
\begin{equation}\label{eq: D1_action}
S_{Dk} = \frac{T_{Dk}}{2} \int d^{k+1} \xi \sqrt{-\hat{g}}\, \bar{\Psi}\, \mathcal{P}_{-} \left[ \slashed{D}_{Dk}+\frac{e^{\phi}}{8\cdot 3!}
F_{ABC}\left(\Gamma^{\hat{M}}\Gamma^{ABC}\Gamma_{\hat{M}}+2\, \Gamma^{ABC}\right)-\frac{1}{2}\Gamma^M \partial_M \phi \right] \Psi\,.
\end{equation}

Varying \cref{eq: D1_action} with respect to the conjugate spinor $\bar\Psi$, we readily obtain the equation of motion of the spinor $\Psi$, 
namely it is
\eq$\label{eq: D1eqmot}
\slashed{D}_{Dk}\Psi+\frac{e^{\phi}}{8\cdot 3!}F_{ABC}\left(\Gamma^{\hat{M}}\Gamma^{ABC}\Gamma_{\hat{M}}+2\, \Gamma^{ABC}\right)
\Psi-\frac{1}{2}(\Gamma^M \partial_M \phi)\Psi=0\,,$ 
where the dilaton field is determined from \cref{eq: dil&RRpot} with $p=1$, thus
\eq$\label{eq: dil1}
e^{\phi}=\left(\frac{R}{r}\right)^{3}\,.$
As it is depicted in \Cref{table: D1_brane_intersections}, there are two possibilities for the probe D$k$-branes. We can either have $k=3$ 
or $k=5$, with $d=0$ and $d=1$, respectively. The R-R field strength field strength in the D$1$/D$k$-brane setup is a 3-form and it is 
given by
\eq$\label{RR1}
F_{(3)}=dC_{(2)}=\frac{6r^{5/2}}{R^{9/2}}\, e^{(0)}\wedge e^{(1)}\wedge\left(\rho\, e^{(\rho)}+\sum_{i=1}^{8-k+d} w^i\,e^{(w^{i})}\right)\,.$
Having in our disposal the components of the R-R field strength as well as the expression of the dilaton field it is straightforward to calculate that
\eq$\label{eq: gammaterm1}
\frac{e^{\phi}}{8\cdot 3!}F_{ABC}\left(\Gamma^{\hat{M}}\Gamma^{ABC}\Gamma_{\hat{M}}+2\, \Gamma^{ABC}\right)\Psi=
\frac{3}{2R^{3/2}}\frac{\rho}{\sqrt{r}} \,\Gamma^{01\rho}\Psi=-\frac{3}{2R^{3/2}}\frac{\rho}{\sqrt{r}} \,\Gamma_{01\rho}\Psi\,,$
while by setting $p=1$ in \cref{eq: dirac&dilterm} we get
\bal$\label{eq: dirac&dil1}
\slashed{D}_{Dk}\Psi^{\ell^{\pm}}-\frac{\Gamma^M\pa_M\phi}{2}\, \Psi^{\ell^{\pm}}=&\left(\frac{R}{r}\right)^{\frac{3}{2}}\Gamma^\mu\pa_\mu
\Psi^{\ell^{\pm}}+\left(\frac{r}{R}\right)^{\frac{3}{2}}\Gamma^\rho\pa_\rho\Psi^{\ell^{\pm}} \pm\frac{1}{\rho}\left(\frac{r}{R}\right)^{\frac{3}{2}} 
\left(\ell + \frac{m}{2} \right)\Psi^{\ell^{\pm}}\nonum\\[2mm]
&+\frac{1}{R^{\frac{3}{2}}}\left(\frac{3}{4}\frac{\rho}{\sqrt{r}}+\frac{m}{2}\frac{r^{\frac{3}{2}}}{\rho}\right)\Gamma^\rho
\Psi^{\ell^{\pm}}\,,$
where $m=k-d-1$.
Substituting  eqs. \eqref{eq: gammaterm1}, \eqref{eq: dirac&dil1} in \eqref{eq: D1eqmot}  and imposing the projection $\Gamma_{01\rho}\Psi=-\Psi$, 
we are led to the following first-order equation of motion
\eq$\label{eq: first_order_D1}
\left\{ \left( \frac{R}{r} \right)^{\frac{3}{2}} \Gamma^{\mu} \partial_{\mu} +  \left( \frac{r}{R} \right)^{\frac{3}{2}} \Gamma^{\rho} \partial_{\rho}
+\frac{1}{R^{\frac{3}{2}}} \left[\frac{3}{2} \frac{\rho}{\sqrt{r}} \pm \left( \ell + \frac{m}{2}  \right) \frac{r^{\frac{3}{2}}}{\rho} +
\left( \frac{3}{4} \frac{\rho}{\sqrt{r}} + \frac{m}{2}\frac{r^{\frac{3}{2}}}{\rho} \right) \Gamma^{\rho}\right]\right\}
\Psi^{\ell^{\pm}} = 0\,.$

In what follows we consider that $\sum_{i} (w^i)^2=L^2$, which by its turn leads to $r^2=\rho^2+L^2$. In both cases, $L$ constitutes the distance 
between the D$1$- and the D$k$-branes. Applying the procedure which was described in \cref{sec: general idea} to 
\cref{eq: first_order_D1}, we construct the following second-order differential equation
\eq$\label{eq: D1_sec_ord_eq}
\left( \mathcal{A}^{\pm}_{1} \partial_{\varrho}^2 + \mathcal{A}^{\pm}_{2} \partial_{\varrho} + \mathcal{A}^{\pm}_{3} \bar{M}^2 +
\mathcal{A}^{\pm}_{4} \Gamma^{\varrho} + \mathcal{A}^{\pm}_{5} \right) \psi^{\ell^{\pm}}(\varrho) = 0\,,$ 
where
\begin{equation} \label{D1: second_order_eom}
\begin{split}
\mathcal{A}^{\pm}_1 &= 1, \quad \mathcal{A}^{\pm}_2 = \frac{m}{\varrho} + \frac{9}{2} \frac{\varrho}{\varrho^2 +1}, \quad \mathcal{A}^{\pm}_3 = 
\frac{1}{\left(\varrho^2 +1 \right)^3}\,,\\
\mathcal{A}^{\pm}_4 &= \frac{3}{2} \frac{1\pm 2 \ell \pm m}{\varrho^2+1} \mp \frac{1}{\varrho^2} \left( \ell + \frac{m}{2} \right) + \frac{3}{2} 
\frac{\varrho^2}{\left( \varrho^2 + 1 \right)^2}\,, \\
\mathcal{A}^{\pm}_5 &= \frac{3}{4} \frac{1}{\varrho^2+1} \left( 3m \mp 2m + 1 \mp 4 \ell \right) - \frac{15}{16} \frac{\varrho^2}{(\varrho^2+1)^2}
 - \frac{1}{\varrho^2} \left( \ell^2 + m \ell+\frac{m}{2} \right)\,.
\end{split}
\end{equation}
Employing the projection $\Gamma^\varrho\psi^{\ell^{\pm}}=\psi^{\ell^{\pm}}$ and adopting the same notation as in \cref{sec: D0} we are led to
\bal$\label{eq: pos_sph_pos_proj_1}
\left[\pa^2_\varrho+\left(\frac{m}{\varrho}+\frac{9}{2}\frac{\varrho}{\varrho^2+1}\right)\pa_\varrho+\frac{\bar{M}^2}{(\varrho^2+1)^{3}}\right.&-
\frac{1}{\varrho^2}\left(\ell^2+(m+1)\ell+m\right)\nonum\\
&\left.+\frac{9}{4}\frac{1+m}{\varrho^2+1}+\frac{9}{16}\frac{\varrho^2}{(\varrho^2+1)^2}\right]\psi^{\oplus}_{\calG_\ell}(\varrho) = 0\,,$
\bal$\label{eq: neg_sph_pos_proj_1}
\left[\pa^2_\varrho+\left(\frac{m}{\varrho}+\frac{9}{2}\frac{\varrho}{\varrho^2+1}\right)\pa_\varrho+\frac{\bar{M}^2}{(\varrho^2+1)^{3}}\right.&-
\frac{1}{\varrho^2}\left(\ell^2+(m-1)\ell\right)\nonum\\
&\left.+\frac{9}{4}\frac{1+m}{\varrho^2+1}+\frac{9}{16}\frac{\varrho^2}{(\varrho^2+1)^2}\right]\psi^{\oplus}_{\calF_\ell}(\varrho) = 0\,,$
for positive and negative spinorial harmonics, respectively. As we already mentioned, $m=k-d-1$, which by its turn implies that for $k=5$ we have
$m=3$, while for $k=3$ we have $m=2$. The corresponding bosonic equations of motion for the D$1$/D$5$ and D$1$/D$3$-brane systems can be 
determined from eqs. \eqref{eq: bosonic_spectra_4} and \eqref{eq: bosonic_spectra_2}, respectively. Hence, setting $p=1$ in the aforementioned 
relations, we get
\eq$\label{eq: bos_eq_15}
\partial^2_{\varrho} f_{\ell}(\varrho) + \frac{3}{\varrho} \partial_{\varrho} f_{\ell}(\varrho) + \left( \frac{\bar{M}^2}{(1 + \varrho^2)^{3}} - 
\frac{\ell(\ell+2)}{\varrho^2} \right) f_{\ell}(\varrho) = 0\,,$
\eq$\label{eq: bos_eq_13}
\partial^2_{\varrho} f_{\ell}(\varrho) + \frac{2}{\varrho} \partial_{\varrho} f_{\ell}(\varrho) + \left( \frac{\bar{M}^2}{(1 + \varrho^2)^{(7-p)/2}} - 
\frac{\ell(\ell+1)}{\varrho^2} \right) f_{\ell}(\varrho) = 0\,,$
respectively. One can easily verify that the transformation
\eq$\label{eq: pos_sph_trans_1}
\psi^{\oplus}_{\calG_{\ell}}(\varrho)=\frac{1}{(1+\varrho^2)^{9/8}}\, f_{\ell+1}(\varrho)\,,$ 
maps \cref{eq: pos_sph_pos_proj_1} either to \cref{eq: bos_eq_15}, when we have the D$1$/D$5$ setup, or to \cref{eq: bos_eq_13},
when we have the D$1$/D$3$ setup. In exactly the same sense, the transformation
\eq$\label{eq: neg_sph_trans_1}
\psi^{\oplus}_{\calF_{\ell}}(\varrho)=\frac{1}{(1+\varrho^2)^{9/8}}\, f_\ell(\varrho)\,,$
maps \cref{eq: neg_sph_pos_proj_1} either to \cref{eq: bos_eq_15} (D$1$/D$5$ setup), or to \cref{eq: bos_eq_13} (D$1$/D$3$ setup). 
The differential equations which result from \eqref{eq: D1_sec_ord_eq} with the projection $\Gamma^\varrho\psi^{\ell^{\pm}}=-\psi^{\ell^{\pm}}$, 
namely the differential equations for the $\psi^\ominus$ spinor, cannot be mapped to the corresponding bosonic equations of motion, although they 
provide us with the same mass eigenvalues $\bar{M}$. The reason behind this is the degeneracy which results from the projection matrix 
$\Gamma^\rho$. Notice as well that the above transformations can be obtained from \eqref{eq: basic_1} by setting $p=1$.

The above equations of motion cannot be solved analytically (as in the case of bosonic mesons) and we have to resort to numerical analysis. We show 
explicitly the first few numerical values of the masses in \Cref{tab: d1}. The numerical values for $n$ and $\ell$ describe the $\mathcal{F}$ operators 
and one needs to shift appropriately the $\ell$ to read off the value of the $\mathcal{G}$ states.
\begin{table}[H]
    \centering
   \begin{tabular}{|c|c|c|c|c|c|}
   \cline{3-6} 
    \multicolumn{2}{c}{}     & \multicolumn{4}{|c|}{$\boldsymbol{n}$} \\ \cline{3-6}
    \multicolumn{2}{c|}{}        & 0 & 1 & 2 & 3       \\ \cline{1-6}
    \multirow{4}{0.5em}{$\boldsymbol{\ell}$}  
    						  & 0 & 14.830 & 49.711 & 104.53 & 179.30 \\ \cline{2-6}  
                             & 1 & 42.793 & 92.952 & 162.67 & 252.03 \\ \cline{2-6}
                             & 2 & 84.366 & 150.11 & 235.27 & 339.96 \\ \cline{2-6}
                             & 3 & 139.43 & 220.76 & 319.85 & 441.55 \\ \cline{1-6}
  \end{tabular}
\hfill%
     \begin{tabular}{|c|c|c|c|c|c|}
     \cline{3-6} 
    \multicolumn{2}{c}{}     & \multicolumn{4}{|c|}{$\boldsymbol{n}$} \\ \cline{3-6}
    \multicolumn{2}{c|}{}        & 0 & 1 & 2 & 3       \\ \cline{1-6}
    \multirow{4}{0.5em}{$\boldsymbol{\ell}$}  
    						  & 0 & 6.2638 	& 34.138 & 82.315 & 150.64 \\ \cline{2-6}  
                             & 1 & 27.082 	& 69.506 & 131.62 & 213.50 \\ \cline{2-6}
                             & 2 & 61.893	& 119.84 & 197.26 & 294.26 \\ \cline{2-6}
                             & 3 & 110.21 	& 183.75 & 276	.58 & 387.88 \\ \cline{1-6}
  \end{tabular}
\caption{Numerical results of the parameter $\bar{M}^2$ for the D$1$/D$k$ brane intersections. The left table pertains to the D$1$/D$5$ and the right one depicts results for the D$1$/D$3$ brane intersection.}
\label{tab: d1}
\end{table}
\section{Open-string fluctuations in the D2-background} \label{sec: D2}
The D$2$-background has been extensively studied in the past in the probe-limit \cite{Myers:2006qr} as well as beyond the quenched approximation 
\cite{Erdmenger:2004dk}. The relevant orientation of the branes in order for the boundary theory to preserve eight supercharges is shown in 
\Cref{table: D2_brane_intersections}.
\begin{table}[H]
\begin{center}
\begin{tabular}{ |c|c|c|c|c|c|c|c|c|c|c|c|} 
 \hline
 &&&&&&&&&&\\[-0.95em] 
									& $x^0$ & $x^1$ & $x^2$ & $x^3$ & $x^4$ & $x^5$ & $x^6$ & $x^7$ & $x^8$ & $x^9$			\\  
 \hline
 Background D$2$-brane 	& --- & --- &--- & $\bullet$ & $\bullet$ & $\bullet$ & $\bullet$ & $\bullet$ & $\bullet$ & $\bullet$		\\ 
 \hline
 Probe D$2$-brane 			& --- & $\bullet$ & $\bullet$ & --- & --- & $\bullet$ & $\bullet$ & $\bullet$ & $\bullet$ & $\bullet$	\\
 \hline
 Probe D$4$-brane 			& --- & --- & $\bullet$ & --- & --- & --- & $\bullet$ & $\bullet$ & $\bullet$ & $\bullet$				\\
 \hline
 Probe D$6$-brane 			& --- & --- & --- & --- & --- & --- & --- & $\bullet$ & $\bullet$ & $\bullet$												\\  
 \hline
\end{tabular}
\caption{The BPS brane intersections. In the above notation --- denotes that a brane extends along that particular direction, while $\bullet$ means that 
the coordinate is transverse to the brane.}
\label{table: D2_brane_intersections}
\end{center}
\end{table}
 
The fermionic action is given by \cite{Marolf:2003vf, Marolf:2003ye, Martucci:2005rb} 
\begin{equation}\label{eq: D2_action}
S_{Dk} = \frac{T_{Dk}}{2} \int d^{k+1} \xi \sqrt{-\hat{g}}\, \bar{\Psi}\, \mathcal{P}_{-} \left[ \slashed{D}_{Dk} + \frac{e^{\phi}}{8 \cdot 4!} 
F_{ABCD} \left( \Gamma^{ABCD} - \Gamma^{\hat{M}} \Gamma^{ABCD} \Gamma_{\hat{M}} \right) - \frac{1}{2} \Gamma^M \partial_M \phi \right] \Psi\,.
\end{equation}

Varying \cref{eq: D2_action} with respect to the conjugate spinor $\bar\Psi$, we readily obtain the equation of motion of the spinor $\Psi$, 
namely it is
\eq$\label{eq: D2eqmot}
\slashed{D}_{Dk}\Psi + \frac{e^{\phi}}{8 \cdot 4!} F_{ABCD} \left( \Gamma^{ABCD} - \Gamma^{\hat{M}} \Gamma^{ABCD} \Gamma_{\hat{M}} \right)\Psi
-\frac{1}{2} (\Gamma^M \partial_M \phi) \Psi=0\,,$
where the dilaton field and the R-R field strength, which in this case constitutes a 4-form, can be determined from \cref{eq: dil&RRpot} with $p=2$, 
thus we have
\eq$\label{eq: dil2}
e^{\phi}=\left(\frac{R}{r}\right)^{\frac{5}{4}}\,,$
\eq$\label{eq: RR2}
F_{(4)}=dC_{(3)}=-\frac{5\sqrt{r}}{R^{5/2}}\, e^{(0)}\wedge e^{(1)}\wedge e^{(2)}\wedge\left(\rho\, e^{(\rho)}+\sum_{i=1}^{7-k+d} w^i\,e^{(w^{i})}
\right)\,,$
respectively. As it is depicted in \Cref{table: D2_brane_intersections}, there are three possibilities for the probe D$k$-branes, $k=\{2,4,6\}$ with 
$d=\{0, 1, 2\}$ respectively. Having in our disposal the components of the R-R field strength as well as the expression of the dilaton field it is straightforward 
to calculate that
\eq$\label{eq: gammaterm2}
\frac{e^{\phi}}{8 \cdot 4!}F_{ABCD}\left(\Gamma^{ABCD} - \Gamma^{\hat{M}} \Gamma^{ABCD} \Gamma_{\hat{M}}\right)\Psi=
\frac{5}{4}\frac{\rho}{R^{5/4}\,r^{3/4}}\Gamma_{012\rho}\Psi\,,$
while by setting $p=2$ in eqs. \eqref{eq: dirac&dilterm} we get
\bal$\label{eq: dirac&dil2}
\slashed{D}_{Dk}\Psi^{\ell^{\pm}}-\frac{\Gamma^M\pa_M\phi}{2}\, \Psi^{\ell^{\pm}}=&\left(\frac{R}{r}\right)^{\frac{5}{4}}\Gamma^\mu\pa_\mu
\Psi^{\ell^{\pm}}+\left(\frac{r}{R}\right)^{\frac{5}{4}}\Gamma^\rho\pa_\rho\Psi^{\ell^{\pm}} \pm\frac{1}{\rho}\left(\frac{r}{R}\right)^{\frac{5}{4}} 
\left(\ell + \frac{m}{2} \right)\Psi^{\ell^{\pm}}\nonum\\[2mm]
&+\frac{1}{R^{\frac{5}{4}}}\left(\frac{5}{8}\frac{\rho}{r^{\frac{3}{4}}}+\frac{m}{2}\frac{r^{\frac{5}{4}}}{\rho}\right)\Gamma^\rho\Psi^{\ell^{\pm}}\,.$
Substituting  eqs. \eqref{eq: gammaterm2}, \eqref{eq: dirac&dil2} in \eqref{eq: D2eqmot}  and imposing the projection  $\Gamma_{012\rho}\Psi=\Psi$, 
we are led to the following first-order equation of motion
\eq$\label{eq: first_order_D2}
\left\{ \left( \frac{R}{r} \right)^{\frac{5}{4}} \Gamma^{\mu} \partial_{\mu} +  \left( \frac{r}{R} \right)^{\frac{5}{4}} \Gamma^{\rho} \partial_{\rho} 
+\frac{1}{R^{\frac{5}{4}}} \left[\frac{5}{4} \frac{\rho}{r^{\frac{3}{4}}} \pm \left( \ell + \frac{m}{2} \right) \frac{r^{\frac{5}{4}}}{\rho} \right] +
\frac{1}{R^{\frac{5}{4}}}\left( \frac{5}{8} \frac{\rho}{r^{\frac{3}{4}}} + \frac{m}{2}\frac{r^{\frac{5}{4}}}{\rho} \right) \Gamma^{\rho}\right\}
\Psi^{\ell^{\pm}} = 0\,.$

Applying the procedure which was described in \cref{sec: general idea} to 
\cref{eq: first_order_D2}, we construct the following second-order differential equation
\eq$\label{eq: D2_sec_ord_eq}
\left( \mathcal{A}^{\pm}_{1} \partial_{\varrho}^2 + \mathcal{A}^{\pm}_{2} \partial_{\varrho} + \mathcal{A}^{\pm}_{3} \bar{M}^2 +
\mathcal{A}^{\pm}_{4} \Gamma^{\varrho} + \mathcal{A}^{\pm}_{5} \right) \psi^{\ell^{\pm}}(\varrho) = 0\,,$ 
where
\begin{equation} \label{D2: second_order_eom}
\begin{split}
\mathcal{A}^{\pm}_1 &= 1, \quad \mathcal{A}^{\pm}_2 = \frac{m}{\varrho} + \frac{15}{4} \frac{\varrho}{\varrho^2 +1}, \quad \mathcal{A}^{\pm}_3 = 
\frac{1}{\left(\varrho^2 +1 \right)^{5/2}}\,, \\
\mathcal{A}^{\pm}_4 &= \frac{5}{4} \frac{1\pm 2 \ell \pm m}{\varrho^2+1} \mp \frac{1}{\varrho^2} \left( \ell + \frac{m}{2} \right) + \frac{5}{8} 
\frac{\varrho^2}{\left( \varrho^2 + 1 \right)^2}\,, \\
\mathcal{A}^{\pm}_5 &= \frac{5}{8} \frac{1}{\varrho^2+1} \left( 3m \mp 2m + 1 \mp 4 \ell \right) - \frac{55}{64} \frac{\varrho^2}{(\varrho^2+1)^2}
 - \frac{1}{\varrho^2} \left( \ell^2 + m \ell+\frac{m}{2} \right)\,.
\end{split}
\end{equation}
Employing the projection $\Gamma^\rho\, \psi^{\ell^{\pm}}=\psi^{\ell^{\pm}}$ and using eqs. \eqref{eq: D2_sec_ord_eq} and 
\eqref{D2: second_order_eom} one can readily obtain the fermionic second-order ordinary differential equation for $\psi^{\ell^{\pm}}(\varrho)$. 
In case of positive spinorial harmonics ($\ell^+$), we get the differential equation of $\psi^{\oplus}_{\calG_{\ell}}(\varrho)$, while for negative 
spinorial harmonics ($\ell^-$), the differential equation of $\psi^{\oplus}_{\calF_{\ell}}(\varrho)$ is obtained. The corresponding bosonic equations of 
motion for the D$2$/D$k$-brane setups ($k=\{2,4,6\}$), can be determined from eqs. \eqref{eq: bosonic_spectra_4}-\eqref{eq: bosonic_spectra_0} by 
setting $p=2$. For all possible D$2$/D$k$-brane systems the transformation 
\eq$\label{eq: pos_spher_trans_2}
\psi^{\oplus}_{\calG_{\ell}}(\varrho)=\frac{1}{(1+\varrho^2)^{15/16}}\, f_{\ell+1}(\varrho)\,,$ 
maps the fermionic function $\psi^{\oplus}_{\calG_{\ell}}(\varrho)$ to the bosonic function $f_{\ell+1}(\varrho)$,  while the transformation
\eq$\label{eq: neg_spher_trans_2}
\psi^{\oplus}_{\calF_{\ell}}(\varrho)=\frac{1}{(1+\varrho^2)^{15/16}}\, f_{\ell}(\varrho)\,,$
maps the fermionic function $\psi^{\oplus}_{\calF_{\ell}}(\varrho)$ to the bosonic function $f_{\ell}(\varrho)$. Notice as well that the above 
transformations can be obtained from \eqref{eq: basic_1} by setting $p=2$. 

The above equations of motion cannot be solved analytically (as in the case of bosonic mesons). We show 
explicitly the first few numerical values of the masses in \Cref{tab: d2}. The numerical values for $n$ and $\ell$ describe the $\mathcal{F}$ operators and 
one needs to shift appropriately the $\ell$ to read off the value of the $\mathcal{G}$ states. It is important to stress at this point that, in cases
where the dimensionality of the background and the probe brane is the same, the quantum number $\ell$---contrary to the rest of the cases---should be greater
than zero as dictated by the normalizability of the mode solution in the UV; observe for example the bottom table in \Cref{tab: d2}.
\begin{table}[H]
    \centering
   \begin{tabular}{|c|c|c|c|c|c|} 
   \cline{3-6}
    \multicolumn{2}{c}{}     & \multicolumn{4}{|c|}{$\boldsymbol{n}$} \\ \cline{3-6}
    \multicolumn{2}{c|}{}        & 0 & 1 & 2 & 3       \\ \cline{1-6}
    \multirow{4}{0.5em}{$\boldsymbol{\ell}$}  
    						  & 0 & 11.509 & 37.185 & 76.982 & 130.93 \\ \cline{2-6}  
                             & 1 & 33.403 & 70.419 & 121.15 & 185.66 \\ \cline{2-6}
                             & 2 & 66.197 & 114.96 & 177.34 & 253.38 \\ \cline{2-6}
                             & 3 & 109.75 & 170.30 & 244.10 & 332.17 \\ \cline{1-6}
  \end{tabular}
\hfill%
     \begin{tabular}{|c|c|c|c|c|c|} 
     \cline{3-6}
    \multicolumn{2}{c}{}     & \multicolumn{4}{|c|}{$\boldsymbol{n}$} \\ \cline{3-6}
    \multicolumn{2}{c|}{}        & 0 & 1 & 2 & 3       \\ \cline{1-6}
    \multirow{4}{0.5em}{$\boldsymbol{\ell}$}  
    						  & 0 & 4.9024 & 25.519 & 60.595 & 110.03 \\ \cline{2-6}  
                             & 1 &21.065 & 52.295 & 97.385 & 156.40 \\ \cline{2-6}
                             & 2 &48.453 & 91.334 & 147.86 & 218.09 \\ \cline{2-6}
                             & 3 &86.630 & 141.29 & 209.54 & 291.40 \\ \cline{1-6}
  \end{tabular}
\newline
\vspace*{1 cm}
\newline
    \begin{tabular}{|c|c|c|c|c|c|} 
    \cline{3-6}
    \multicolumn{2}{c}{}     & \multicolumn{4}{|c|}{$\boldsymbol{n}$} \\ \cline{3-6}
    \multicolumn{2}{c|}{}        & 0 & 1 & 2 & 3       \\ \cline{1-6}
    \multirow{4}{0.5em}{$\boldsymbol{\ell}$}  
    						  & 1 & 	11.509 & 37.185 & 76.982 & 130.93 \\ \cline{2-6}  
                             & 2 &	33.403 & 70.149 & 121.15 & 185.66 \\ \cline{2-6}
                             & 3 &	66.197 & 114.96 & 177.34 & 253.38 \\ \cline{1-6}
  \end{tabular}
\caption{Numerical results of the parameter $\bar{M}^2$ for the D$2$/D$k$ brane intersections. The top left table pertains to the 
D$2$/D$6$ and the top right one depicts results for the D$2$/D$4$ brane intersection. The bottom table is related to the D$2$/D$2$ setup.}
\label{tab: d2}
\end{table}
\section{Probing the D3-background} \label{sec: D3}
Before we start the analysis of the fermionic modes for the background generated by a stack of $N_c$ branes we would like to illustrate the brane 
intersections that we examine in this work. This is shown in \Cref{table: D3_brane_intersections}.

\begin{table*}[h]
\begin{center}
		\begin{adjustbox}{max width=\textwidth}
\begin{tabular}{ |c|c|c|c|c|c|c|c|c|c|c|c|}
 \hline
 &&&&&&&&&&\\[-0.95em] 
 					  					& $x^0$ & $x^1$ & $x^2$ & $x^3$ & $x^4$ & $x^5$ & $x^6$ & $x^7$ & $x^8$ & $x^9$			\\ 
 \hline
 Background D$3$-brane 	& --- & --- & --- & --- & $\bullet$ & $\bullet$ & $\bullet$ & $\bullet$ & $\bullet$ & $\bullet$		\\ 
 \hline
 Probe D$3$-brane 			& --- & --- & $\bullet$ & $\bullet$ & --- & --- & $\bullet$ & $\bullet$ & $\bullet$ & $\bullet$				\\
 \hline
 Probe D$5$-brane 			& --- & --- & --- & $\bullet$ & --- & --- & --- & $\bullet$ & $\bullet$ & $\bullet$				\\
 \hline
 Probe D$7$-brane 			& --- & --- & --- & --- & --- & --- & --- & --- & $\bullet$ & $\bullet$												\\  
 \hline
\end{tabular}
		\end{adjustbox}
\caption{The brane intersections that preserve eight supercharges. In the above notation --- denotes that a brane extends along that particular direction, 
while $\bullet$ means that the coordinate is transverse to the brane.}
\label{table: D3_brane_intersections}
\end{center}
\end{table*}
The fermionic action in the D$3$-brane background is \cite{Marolf:2003vf, Marolf:2003ye, Martucci:2005rb}
\begin{equation}\label{eq: D3_action}
S_{Dk} = \frac{T_{Dk}}{2} \int d^{k+1} \xi \sqrt{-\hat{g}}\, \bar{\Psi}\, \mathcal{P}_{-} \left( \slashed{D}_{Dk} + \frac{i}{2 \cdot 8 \cdot 5!} 
\Gamma^{\hat{M}} F_{ABCDE}\, \Gamma^{ABCDE}\,  \Gamma_{\hat{M}} \right) \Psi.
\end{equation}

Varying \cref{eq: D3_action} with respect to the conjugate spinor $\bar\Psi$, we readily obtain the equation of motion of the spinor $\Psi$, 
namely it is
\eq$\label{eq: D3eqmot}
\slashed{D}_{Dk}\Psi + \frac{i}{2 \cdot 8 \cdot 5!} \Gamma^{\hat{M}} F_{ABCDE}\, \Gamma^{ABCDE}\, \Gamma_{\hat{M}}\Psi=0\,.$
In this particular scenario the dilaton field vanishes, while the R-R field strength is a self-dual 5-form, which is determined with the use of 
\cref{eq: dil&RRpot} for $p=3$, namely we have \footnote{For more details about the Hodge star operator $\star$ see 
\Cref{app: conventions}.}
\eq$\label{eq: RR3}
F_{(5)}=dC_{(4)}+\star(\displaystyle{dC_{(4)}})\,.$
As it is depicted in \Cref{table: D3_brane_intersections}, there are three possibilities for the probe D$k$-branes, $k=\{3,5,7\}$ with 
$d=\{1, 2, 3\}$ respectively. In case of the D$3$/D$3$-brane setup, the above relation leads to
\bal$\label{eq: RR33}
F_{(5)}=\frac{4}{rR}&\left[e^{(0)}\wedge e^{(1)}\wedge e^{(2)}\wedge e^{(3)}\wedge\left(\rho\, e^{(\rho)}+\sum_{m=1}^{4} w^m\,  e^{(w^m)}\right)
-\rho\, e^{(\theta)}\wedge e^{(w^{1})}\wedge e^{(w^{2})}\wedge e^{(w^{3})}\wedge e^{(w^{4})}\right.\nonum\\[2mm]
&-e^{(\rho)}\wedge e^{(\theta)}\wedge\left(w^1\, e^{(w^{2})}\wedge e^{(w^{3})}\wedge e^{(w^{4})}-w^2\, e^{(w^{1})}\wedge e^{(w^{3})}\wedge 
e^{(w^{4})}+w^3\, e^{(w^{1})}\wedge e^{(w^{2})}\wedge e^{(w^{4})}\right.\nonum\\[2mm]
&\left.\left.-w^4\, e^{(w^{1})}\wedge e^{(w^{2})}\wedge e^{(w^{3})}\right)\right]\,.$
In case of the D$3$/D$5$-brane setup, it is
\bal$\label{eq: RR35}
F_{(5)}=\frac{4}{rR}&\left[e^{(0)}\wedge e^{(1)}\wedge e^{(2)}\wedge e^{(3)}\wedge\left(\rho\, e^{(\rho)}+\sum_{m=1}^{3} w^m\,  e^{(w^m)}\right)
-\rho\, e^{(S^{2})}\wedge e^{(w^{1})}\wedge e^{(w^{2})}\wedge e^{(w^{3})}\right.\nonum\\[2mm]
&\left.+e^{(\rho)}\wedge e^{(S^{2})}\wedge\left(w^1\, e^{(w^{2})}\wedge e^{(w^{3})}-w^2\, e^{(w^{1})}\wedge e^{(w^{3})}+w^3\, e^{(w^{1})}\wedge 
e^{(w^{2})}\right)\right]\,,$
where $e^{(S^{2})}=e^{(\theta_1)}\wedge e^{(\theta_2)}$. Finally, in case of the D$3$/D$7$-brane setup, we have
\bal$\label{eq: RR37}
F_{(5)}=\frac{4}{rR}&\left[e^{(0)}\wedge e^{(1)}\wedge e^{(2)}\wedge e^{(3)}\wedge\left(\rho\, e^{(\rho)}+\sum_{m=1}^{2} w^m\,  e^{(w^m)}\right)
-\rho\, e^{(S^{3})}\wedge e^{(w^{1})}\wedge e^{(w^{2})}\right.\nonum\\[2mm]
&\left.-e^{(\rho)}\wedge e^{(S^{3})}\wedge\left(w^1\, e^{(w^{2})}-w^2\, e^{(w^{1})}\right)\right]\,,$
where $e^{(S^{3})}=e^{(\theta_1)}\wedge e^{(\theta_2)}\wedge e^{(\theta_3)}$.
Having in our disposal the components of the R-R field strength as well as the expression of the dilaton field it is straightforward 
to calculate that regardless the D$3$/D$k$-brane system it is
\eq$\label{eq: gammaterm3}
\frac{i}{2 \cdot 8 \cdot 5!} \Gamma^{\hat{M}} F_{ABCDE}\, \Gamma^{ABCDE}\, \Gamma_{\hat{M}}\Psi=i\,\frac{\rho}{rR}\Gamma_{0123\rho}\Psi\,,$
while by setting $p=3$ in \eqref{eq: dirac&dilterm} we get
\bal$\label{eq: dirac&dil3}
\slashed{D}_{Dk}\Psi^{\ell^{\pm}}-\frac{\Gamma^M\pa_M\phi}{2}\, \Psi^{\ell^{\pm}}=\left\{\frac{R}{r}\,\Gamma^\mu\pa_\mu
+\frac{r}{R}\,\Gamma^\rho\pa_\rho\pm\frac{1}{\rho}\frac{r}{R}\left(\ell + \frac{m}{2} \right)
+\frac{1}{R}\left(\frac{1}{2}\frac{\rho}{r}+\frac{m}{2}\frac{r}{\rho}\right)\Gamma^\rho\right\}\Psi^{\ell^{\pm}}\,.$
Substituting  eqs. \eqref{eq: gammaterm3}, \eqref{eq: dirac&dil3} in \eqref{eq: D3eqmot}  and imposing the projection  $\Gamma_{0123\rho}\Psi=-i
\Psi$, we are led to the following first-order equation of motion
\eq$\label{eq: first_order_D3}
\left\{ \frac{R}{r} \Gamma^{\mu} \partial_{\mu} +   \frac{r}{R}  \Gamma^{\rho} \partial_{\rho} +\frac{1}{R} \left[ \frac{\rho}{r} \pm \left( \ell + 
\frac{m}{2} \right) \frac{r}{\rho} \right] +\frac{1}{2R}\left( \frac{\rho}{r} + m\,\frac{r}{\rho} \right) \Gamma^{\rho}\right\}
\Psi^{\ell^{\pm}} = 0\,.$

Applying the procedure which was described in \cref{sec: general idea} to 
\cref{eq: first_order_D3}, we construct the following second-order differential equation
\eq$\label{eq: D3_sec_ord_eq}
\left( \mathcal{A}^{\pm}_{1} \partial_{\varrho}^2 + \mathcal{A}^{\pm}_{2} \partial_{\varrho} + \mathcal{A}^{\pm}_{3} \bar{M}^2 +
\mathcal{A}^{\pm}_{4} \Gamma^{\varrho} + \mathcal{A}^{\pm}_{5} \right) \psi^{\ell^{\pm}}(\varrho) = 0\,,$ 
where
\begin{equation} \label{D3: second_order_eom}
\begin{split}
\mathcal{A}^{\pm}_1 &= 1, \quad \mathcal{A}^{\pm}_2 = \frac{m}{\varrho} + \frac{3 \varrho}{\varrho^2 +1}, \quad \mathcal{A}^{\pm}_3
 = \frac{1}{\left(\varrho^2 +1 \right)^2}\,, \\
\mathcal{A}^{\pm}_4 &= \frac{1\pm 2 \ell \pm m}{\varrho^2+1} \mp \frac{1}{\varrho^2} \left( \ell + \frac{m}{2} \right)\,,  \\
\mathcal{A}^{\pm}_5 &= \frac{1}{\varrho^2+1} \left( \frac{3m+1\mp 2m}{2} \mp 2 \ell \right) - \frac{3}{4} \frac{\varrho^2}{(\varrho^2+1)^2}
 - \frac{1}{\varrho^2} \left( \ell^2 + m \ell+\frac{m}{2} \right)\,.
\end{split}
\end{equation}
Employing the projection $\Gamma^\rho\, \psi^{\ell^{\pm}}=\psi^{\ell^{\pm}}$ and using eqs. \eqref{eq: D3_sec_ord_eq} and 
\eqref{D3: second_order_eom} one can readily obtain the fermionic second-order ordinary differential equation for $\psi^{\ell^{\pm}}(\varrho)$. 
In case of positive spinorial harmonics ($\ell^+$), we get the differential equation of $\psi^{\oplus}_{\calG_{\ell}}(\varrho)$, while for negative 
spinorial harmonics ($\ell^-$), the differential equation of $\psi^{\oplus}_{\calF_{\ell}}(\varrho)$ is obtained. The corresponding bosonic equations of 
motion for the D$3$/D$k$-brane setups ($k=\{3,5,7\}$), can be determined from eqs. \eqref{eq: bosonic_spectra_4}-\eqref{eq: bosonic_spectra_0} by 
setting $p=3$. For all possible D$3$/D$k$-brane systems the transformation 
\eq$\label{eq: pos_spher_trans_3}
\psi^{\oplus}_{\calG_{\ell}}(\varrho)=\frac{1}{(1+\varrho^2)^{3/4}}\, f_{\ell+1}(\varrho)\,,$ 
maps the fermionic function $\psi^{\oplus}_{\calG_{\ell}}(\varrho)$ to the bosonic function $f_{\ell+1}(\varrho)$,  while the transformation
\eq$\label{eq: neg_spher_trans_3}
\psi^{\oplus}_{\calF_{\ell}}(\varrho)=\frac{1}{(1+\varrho^2)^{3/4}}\, f_{\ell}(\varrho)\,,$
maps the fermionic function $\psi^{\oplus}_{\calF_{\ell}}(\varrho)$ to the bosonic function $f_{\ell}(\varrho)$. Notice as well that the above 
transformations can be obtained from \eqref{eq: basic_1} by setting $p=3$.
	\subsection{Analytic solutions} \label{sec: D3_analytics}
The equation of motion which result from \eqref{D3: second_order_eom} admit analytic solutions. Although they have been previously derived 
in \cite{Kirsch:2006he,Abt:2019tas} for massless and massive flavours respectively, we present them here as well for completeness.  The solution of the
differential equation related to the positive spinor spherical harmonics is of the following form
\begin{equation} \label{eq: solutions_D3_positive_harmonics}
\begin{split}
\psi_{\mathcal{G}} = ~ &\frac{\varrho^{\ell+1}}{(1 + \varrho^2)^{\left(n+\ell+\frac{m}{2} +\frac{5}{4} \right)}} ~ {_2}F_1 \left( -n, -n-\ell-\frac{1}{2} (m+1); 
\ell + \frac{m+3}{2}; - \varrho^2  \right) \chi_{+}  \\
& +\frac{\varrho^{\ell}}{(1 + \varrho^2)^{\left(n+\ell+\frac{m}{2} +\frac{5}{4} \right)}} ~ {_2}F_1 \left( -n, -n-\ell-\frac{1}{2}(m+3);\ell+\frac{m+1}{2}; - 
\varrho^2  \right) \chi_{-}\,.
\end{split}
\end{equation}
In the above, the number $n$ is directly related to the mass spectrum via the relation
\begin{align}
 \bar{M}^2_{\mathcal{G}} &= 4 \left(n + \ell + \frac{m+1}{2} \right) \left(n + \ell + \frac{m+3}{2}  \right)\,,
\end{align}
where $n,\ \ell\geq 0$. The spinors $\chi_{\pm}$ are eigenstates of the $\Gamma^\varrho$-matrix and they obey the relation $\Gamma^{\varrho} 
\chi_{\pm} = \pm \chi_{\pm}$. They are related to one another via 
\begin{equation}
\chi_{-} = \frac{i \slashed{k}}{M} \chi_{+}\,,
\end{equation}
where $k$ is the wave-vector from the plane-wave ansatz in the decomposition of the ten-dimensional spinor. The asymptotic behaviour of the above solution
in the UV ($\varrho \rightarrow \infty$) and the IR ($\varrho \rightarrow 0$) are given below
\begin{equation} \label{eq: asymptotics_positive_states}
\begin{split}
\psi_{\mathcal{G}}|_{\varrho \rightarrow \infty} &\sim \varrho^{-\ell-m-3/2}  \chi_{+} + \varrho^{-\ell-m-5/2}  \chi_{-}\,, \\
\psi_{\mathcal{G}}|_{\varrho \rightarrow 0} &\sim \varrho^{\ell+1} ~ \chi_{+} + \varrho^{\ell} ~ \chi_{-}\,.
\end{split}
\end{equation}

We can perform the same analysis and derive the solutions describing the $\mathcal{F}$, the ones obtained by considering the negative eigenvalues on the 
internal manifold.	 They are 
\begin{equation} \label{eq: solutions_D3_negative_harmonics}
\begin{split}
\psi_{\mathcal{F}} = ~ &\frac{\varrho^{\ell}}{(1 + \varrho^2)^{\left(n+\ell+\frac{m}{2} +\frac{1}{4} \right)}} ~ {_2}F_1 \left( -n, -n-\ell-\frac{1}{2} (m-1); 
\ell + \frac{m+1}{2}; - \varrho^2  \right) \chi_{+}  \\
&+ \frac{\varrho^{\ell+1}}{(1 + \varrho^2)^{\left(n+\ell+\frac{m}{2} +\frac{1}{4} \right)}} ~ {_2}F_1 \left( -n, -n-\ell-\frac{1}{2}(m+1);\ell+\frac{m+3}{2}; 
- \varrho^2  \right) \chi_{-}\,,
\end{split}
\end{equation}
leading to the discrete mass spectrum
\eq$
 \bar{M}^2_{\mathcal{F}} = 4 \left(n + \ell + \frac{m-1}{2} \right) \left(n + \ell + \frac{m+1}{2}  \right)\,,$
$n$, $\ell \geq 0$. The asymptotic behaviour of the above solution in the UV ($\varrho \rightarrow \infty$) and the IR ($\varrho \rightarrow 0$) are
given below
\begin{equation} \label{eq: asymptotics_negative_states}
\begin{split}
\psi_{\mathcal{F}}|_{\varrho \rightarrow \infty} &\sim \varrho^{-\ell-m-1/2} ~ \chi_{+} + \varrho^{-\ell-m+1/2} ~ \chi_{-}\,, \\
\psi_{\mathcal{F}}|_{\varrho \rightarrow 0} &\sim \varrho^{\ell} ~ \chi_{+} + \varrho^{\ell+1} ~ \chi_{-}\,.
\end{split}
\end{equation}
\subsection{Baryons and the large-$N_c$ limit} \label{sec: D3_largeNc}
In this section we would like to specify our discussion to the case of a four-dimensional strongly coupled gauge theory without defects and hence we will be 
focused on the D$3$/probe-D$7$ intersection. The extension of the analysis to lower dimensional gauge theories based on the D$3$-background geometry is 
straightforward and similar arguments can be given for theories with quarks confined on one and two-dimensional defects. 

We have studied, thus far, the dynamics and solved for the spectra of the fermionic superpartners of mesons---dubbed mesinos---which have no counterpart in 
ordinary non-supersymmetric field theories. For phenomenological applications of holography, however, it is interesting to study fermionic degrees of freedom 
in the bulk. As it was pointed out in \cite{Abt:2019tas} a certain class of these mesino states, namely the $\mathcal{G}$ string modes, resemble structurally the 
baryons of ordinary QCD. This observation was made at the level of the analysis of the fluctuation to operator matching. It was observed that these states are 
comprised out of three elementary fermionic fields, two quarks and a gaugino.  

On the formal side of the analysis, the true dynamical baryonic vertex in the string theory setup has been described in \cite{Witten:1998xy} and it consists of 
our base system with the D$7$-probe in addition to which we add D$5$-branes wrapping a spherical subspace inside the initial $AdS_5 \times S^5$
background geometry and having $N_c$ strings ending on it. One can immediately see the complications. Since in the dynamical baryonic vertex $N_c$ strings
are ending on the baryonic (D$5$-)brane it is unclear whether or not the probe-approximation can be made and what its regime of validity is if any. Also, since
the brane intersection consists of several different branes, the topology of the system of interest is inherently complicated. Finally, baryons are half-integer
fields and their dynamics are governed by the Dirac or the Rarita-Schwinger actions which have to be studied on highly curved manifolds. 

Now, we would like to give additional evidence and support the notion that the $\mathcal{G}$-states (which we remind the reader are $\mathcal{G} \sim
\bar{\psi} \lambda \psi$, where $\bar{\psi},\psi$ are fundamental Fermi fields---quark like states---and $\lambda$ is an adjoint fermion---a gaugino field)
can be thought of as effectively describing baryonic operators. In order to amplify and re-express the argument of \cite{Abt:2019tas} regarding the possible
baryonic interpretation of these supergravity states, we will show that their mass eigenvalues obey the correct large-$N_c$ scaling, dictated by field theory
considerations \cite{Witten:1979kh}. This has already been done in an effective, bottom-up approach in \cite{Kirsch:2006he}. More precisely the case of
massless quarks/overlapping branes was considered in that work.  

We start the analysis of this section by recalling that the scaling dimension and the $AdS_5$ mass for a spin-$1/2$ field is given by \cite{Henningson:1998cd}
\begin{equation} \label{eq: dimension_mass_relation}
\Delta_{\ell} = |m_{\ell}| + 2 \,,
\end{equation}
and for the states considered so far we have $m^{\mathcal{G}}_{\ell}  = \frac{5}{2} + \ell$ and $m^{\mathcal{F}}_{\ell}  = - \left(\frac{1}{2} +\ell \right)$ 
for the $\mathcal{G}$ and $\mathcal{F}$ states respectively. 

Now, let us turn our attention to baryonic operators in a supersymmetric Yang-Mills theory with an $SU(N_c)$ gauge group and a certain number of flavours.
We will be assuming that the theory is such that it is conformal, or perhaps more generally it possesses a regime in its phase space with walking dynamics
(conformal window). There are many examples of such four-dimensional field theories that are asymptotically free, for a discussion on the phase diagram of
supersymmetric $SU(N_c)$ Yang-Mills theories see \cite{Ryttov:2007sr}, but here we just mention the four-dimensional $SU(3)$ super-QCD with $9/2 \leq
N_f \leq 9$ flavours as an example of the theories we mentioned. The existence of the conformal phase of the theory makes sure that the holographic gravity
description has an $AdS_5$ structure. In the aforementioned class of theories a baryon is the colour singlet composite bound state comprised out of $N_c$
quarks, and a baryonic operator is the totally anti-symmetrized object given by 
\begin{equation}
\mathcal{B} \sim  \epsilon^{\overbrace{ijk\cdots y z}^{N_{c}-indices}} ~ \underbrace{{q_i q_j q_k \cdots q_y q_z}}_{N_{c}-fields}\,,
\end{equation}
where we have written the lowest entry of the super(conformal) multiplet up to an irrelevant numerical coefficient which is a function of $N_c$. We can
construct higher baryonic excitations in the multiplet by applying an $\ell$ number of times the derivative operator to the above. The associated conformal
dimension is given by 
\begin{equation} \label{eq: baryon_conformal_dimension}
\Delta = \frac{3}{2} N_c + \ell\,.
\end{equation}
It is straightforward to use \eqref{eq: dimension_mass_relation} and obtain 
\begin{equation} \label{eq: ads_baryon_mass}
m^{\mathcal{B}}_{\ell} = \frac{3}{2} N_c + \ell - 2\,.
\end{equation}

We have already seen that the positive spinor spherical harmonics give rise to states that resemble baryons of non-supersymmetric QCD. We re-write the
eigenvalues of the spinor spherical harmonic in a different but equivalent form for the $\mathcal{G}$-modes
\begin{equation} \label{eq: spinor harmonics_2}
\slashed{D}_{S^{3}}\Psi^{\ell^{+}} = \left(m^{\mathcal{G}}_{\ell}- 1\right)\Psi^{\ell^{+}}\,.
\end{equation}
At this point we replace in the above relation \cref{eq: spinor harmonics_2} the bulk-$AdS$ mass for a baryonic operator given by 
\cref{eq: ads_baryon_mass} and hence we obtain 
\begin{equation} \label{eq: spinor harmonics_3}
\slashed{D}_{S^{3}}\Psi = \left(m^{\mathcal{B}}_{\ell}- 1\right)\Psi\,.
\end{equation}
It is a matter of simple algebra to derive the differential equation by using \cref{eq: spinor harmonics_3} and the mathematical procedure explained 
in the previous section. Thus, we get 
\bal$ \label{eq: baryon_eqn_of_motion}
\Bigg\{ \partial^2_{\varrho} &	+\left[\frac{3\varrho}{1+\varrho^2}+\frac{2(m^{\mathcal{B}}_{\ell}-\ell-1)}{\varrho}\right] \partial_{\varrho} +
\frac{\bar{M}^2_{\mathcal{B}}}{(1+\varrho^2)^2} - \frac{3}{4} \frac{\varrho^2}{(1+\varrho^2)^2}- \frac{1}{\varrho^2} \Big[ m^{\mathcal{B}}_{\ell}
(2\ell+1)-(\ell+1)^2-\ell \nonum\\[1mm]
&+ (m^{\mathcal{B}}_{\ell}-1)\Gamma^{\varrho} \Big] 
+\frac{1}{1+\varrho^2} \left[m^{\mathcal{B}}_{\ell}-3\ell-\frac{1}{2}+(2m^{\mathcal{B}}_{\ell}-1) \Gamma^{\varrho} \right] \Bigg\} \Psi =0\,.$
In order to obtain the solution to the above equation of motion representing the supergravity states we require the solution to be normalizable 
in the UV and regular in the IR, hence, we have
\begin{equation} \label{eq: solutions_baryon_states}
\begin{split}
\psi_{\mathcal{B}} = &\frac{\varrho^{\ell + 1}}{(1 + \varrho^2)^{ n + \ell + \frac{3 N_c}{2}-\frac{7}{4} }} ~ {_2}F_1 \left( -n, - n - 
\ell - \frac{3N_c-5}{2} ; \ell + \frac{3(N_c-1)}{2}; - \varrho^2  \right) \chi_{+} \\
&+\frac{\varrho^{-\frac{9}{2} + \ell + \frac{3N_c}{2}}}{(1 + \varrho^2)^{-\frac{7}{4} + n + \ell + \frac{3 N_c}{2}}} ~ {_2}F_1 \left( -n, - n-\ell-
\frac{3(N_c-1)}{2} ; \ell+\frac{3N_c-5}{2};  - \varrho^2  \right) \chi_{-} \,,
\end{split}
\end{equation}
while the discrete mass spectrum is shown below
\begin{equation} \label{eq: baryon_mass_spectrum}
\bar{M}_{\mathcal{B}} = 2 \sqrt{\left(n + \ell + \frac{3}{2} N_c - \frac{5}{2} \right) \left( n + \ell + \frac{3}{2} (N_c-1) \right)}\,.
\end{equation}
It is obvious that in the $N_c \rightarrow \infty$ limit the mass scales with $N_c$ as desired \cite{Witten:1979kh}. It is quite obvious as well, that 
for $N_c=3$ eqs. \eqref{eq: solutions_baryon_states} and \eqref{eq: baryon_mass_spectrum} reduce to the ones derived in \Cref{sec: D3_analytics} above
for the D3/probe-D7 setup. 

Let us conclude the discussion of this section by stating the basic points. The large-$N_c$ limit scaling of the mass was derived in \cite{Kirsch:2006he} for the
case of overlapping branes in a bottom-up way effectively. Here, we provide a derivation from the top-down construction in a set-up that describes massive
dynamical quarks by construction. We should stress---though obvious---that our result is derived by assuming that the identification of the $\mathcal{G}$
modes as describing baryonic operators at least in some regime of the parameter space of the gauge theory is correct. The precise statement we wish to make
here is the same as the one made in \cite{Kirsch:2006he}. If one realizes and manages to solve the brane configuration that yields a dynamical baryon in the
string theory side with an $AdS$ geometric part, then the result derived in eq. \eqref{eq: baryon_mass_spectrum} should be re-obtained from the full brane
intersection; it should at least hold in a certain regime of the parameter space of the full setup.
\subsection{Double-trace boundary interactions and avoided level crossings} \label{sec: multi_trace}
The models that we have considered thus far cannot give rise to abnormally light fermionic states compared to the scale set by the vector meson mass of the brane intersection. The reason behind this, lies to the residual amount of supersymmetry---after the introduction of flavour branes---which ties these masses to the spectra of the bosonic meson states. It is, however, useful for applications of holographic models to understand if and how one is able to make certain states lighter. A neat example is the study of holographic composite Higgs models \cite{Erdmenger:2020lvq} where the double-trace interactions are necessary to ensure the correct mass for the top. Double-trace deformations are also used in other bottom-up applications in holography. They have been used, for example, to describe the colour superconductivity \cite{BitaghsirFadafan:2018iqr} where the double-trace interactions are used to generate the Cooper pair condensate.

With holographic applications in mind, the authors in \cite{Abt:2019tas} considered double-trace deformations of the boundary Lagrangian of ``mesino'' squared type. The way that this was achieved was to use the prescription for higher dimensional operators described by Witten \cite{Witten:2001ua}. The addition of such operators is
naively a reduction to the fermionic mass in the low-energy hadronic description. It is worthwhile mentioning that the method with the double-trace interactions
has been used in the past in the context of the D$3$/probe-D$7$ brane intersection to describe the Nambu-Jona-Lasinio model \cite{Evans:2016yas}. This
model utilizes four-fermi interactions. 

The precept of the exercise with double-trace operators of ``mesino'' squared type is that any state can be driven much lighter compared to the undeformed 
$\mathcal{N}=2$ theory, however it can never be made lighter than the previous one in the supermultiplet. Hence, the authors in \citep{Abt:2019tas} found an
avoided level-crossing. This was observed only numerically and an analytic explanation is still lacking. Here we provide an analytic explanation. 

The effects of adding these higher dimension operators should manifest themselves as new phenomena at some scale in the UV which we call $\Lambda_{UV}$.
These operators deform the original Lagrangian in the following schematic way: 
\begin{equation}
\mathcal{L} + \frac{g^2}{\Lambda^q_{UV}} ~ \bar{\mathcal{O}} ~ \mathcal{O}\,,
\end{equation}
where the power $q$ is chosen with relevance to the conformal dimension of the operator.  Witten argued that if the operator obtains a non-trivial vacuum
expectation value, it generates a source at the UV boundary, which is given by 
\begin{equation} \label{eq: basic_dt}
\mathcal{J} = \frac{g^2}{\Lambda^q_{UV}} \langle \mathcal{O} \rangle\,.
\end{equation}

All the spectra presented so far, and the the spectra that we present in \Cref{sec: D4} for the D$4$ background, are solutions with a vanishing source; 
$\mathcal{J}=0$. In order to include the double-trace deformation of the Lagrangian, we allow for solutions to have $\mathcal{J} \neq 0$, and from the
asymptotic expansions of these modes in the UV we can read off the precise values of $\mathcal{O}$ and $\mathcal{J}$. The asymptotic expansions have 
been obtained in the past and the precise explanation of the sources and operators in terms of these expansions are given in \citep{Abt:2019tas,Laia:2011wf}
and we do not repeat the analysis here. Then, by using \cref{eq: basic_dt} we are able to obtain the coupling $g^2$. These new solutions that have 
$\mathcal{J} \neq 0$ are interpreted as being part of the sourceless theory in the presence of the higher dimensional operators.

Having explained the way that these higher dimensional operators can be included in our base system, let us focus, without loss of generality, on the 
D$3$/D$7$ brane junction and the modes associated to the $\mathcal{G}$ operators. For a certain value of the angular momentum $\ell=0$ we ask the
question whether or not the first radially excited state of the KK tower, $n=1$, can be driven lighter than the ground state, $n=0$. There is numerical evidence
that this does not happen.

Since we have already argued that the same arguments can be obtained from either of the $\Gamma^{\varrho}$ projections we will be using the $\oplus$ 
states for convenience. Under the change of variables: 
\begin{equation}
\varrho = e^{-y}, \qquad \psi(\varrho) = \frac{e^{-2y}}{(1+e^{2y})^{3/4}} \varphi(y),
\end{equation}
the equations of motion can be brought in a Schr\"odinger form, namely 
\begin{equation}
(\partial^2_y  - V(y)) \varphi(y) = 0,
\end{equation}
with the potential being given by:
\begin{equation}
V(y) = (\ell+2)^2 - \frac{1}{4} \sech^2 y ~ \bar{M}^2\,.
\end{equation}
We drop the explicit $y$-dependence for notational convenience and we denote the ground state ($n=0$, $\ell=0$) by $\varphi_0$ and the
corresponding mass eigenvalue by $M^2_0$. Likewise we use the subscript $1$ for the $n=1,\ \ell=0$ state or any state that lies above the ground 
state ($n>0$, $\ell=0$) of the system and has a higher mass with a non-vanishing source in the UV. Hence, we obtain
\begin{equation}
\begin{split}
\partial^2_y \varphi_0   - V_0 \varphi_0 &= 0\,,\\[2mm]
\partial^2_y \varphi_1   - V_1 \varphi_1 &= 0\,.
\end{split}
\end{equation}
If a level-crossing is to occur it would mean that $\bar{M}^2_0=\bar{M}^2_1$ (and $V_0=V_1$) for some value of a specified source term. We 
multiply the first equation by $ \varphi_1$ and the second by $ \varphi_0$ and we subtract one from the other to obtain 
\gat$
\varphi_1 \, \partial^2_y \varphi_0 - \varphi_0 \, \partial^2_y \varphi_1 = 0 \Ra\nonum \\[1mm]
\partial_y \left( \varphi_1 \, \partial_y \varphi_0 - \varphi_0 \, \partial_y \varphi_1 \right) = 0 \Ra\nonum \\[1mm]
\varphi_1 \, \partial_y \varphi_0 - \varphi_0 \, \partial_y \varphi_1 = c \,,$
where $c$ is just a finite constant. Note that if we are dealing with the normalizable modes only, then we can use the boundary conditions at infinity to set
$c=0$ and we solve the differential equation above to obtain $\varphi_0 = \wtilde{c} ~ \varphi_1$, where now $\wtilde{c}$ is a new constant. However, this
would only mean that there is no deformation of the boundary Lagrangian and we have only the states of the undeformed $\mathcal{N}=2$ theory. If we 
wish to allow for a non-negligible double-trace deformation we should formally write 
\begin{equation} \label{eq:condition_1_schr}
 \varphi_1 \, \partial_y \varphi_0 - \varphi_0 \,\partial_y \varphi_1 = c \,,
\end{equation}
for some finite value of the constant $c$. Note that while we know that $\varphi_0$ is the ground state and hence it falls off to $0$ in the UV we cannot set
the constant to $c=0$ as the term with the non-vanishing source might be going to some finite value faster.

\Cref{eq:condition_1_schr} is a necessary condition for a level-crossing to occur, thus we can re-express it in terms of the $\varrho$ coordinate and 
the spinorial wave-functions $\psi_{(0,1)}$ in the following way:
\begin{equation} \label{eq:condition_2_schr}
\varrho (1+\varrho^2)^{3/2}\, \partial_{\varrho} \left[ \ln \left(\frac{\psi_0}{\psi_1} \right) \right] = \frac{c}{\psi_0 \, \psi_1}.
\end{equation}
We can readily obtain the form of $\psi_0$ by using \cref{eq: solutions_D3_positive_harmonics} for the special values $n=\ell=0$, $m=3$ and taking
into account the positive $\Gamma^{\varrho}$-projection, thus we get
\begin{equation} \label{eq:simplemode}
\psi_0 = \frac{\varrho}{(1+\varrho^2)^{11/4}}.
\end{equation}
However, in order to obtain the form of $\psi_1$ we need to reconsider the solutions of the second-order equations of motion. The difference now is that we
will allow as solutions the more general form and not just the piece that falls off asymptotically to zero as $\varrho \ra\infty$. For our convenience we 
re-state the second-order differential equation of the $\mathcal{G}$-modes for the positive $\Gamma^{\varrho}$-projection:
\begin{equation}
\left[ \partial^2_{\varrho} + \left(\frac{3}{\varrho} + \frac{3 \varrho}{1+\varrho^2} \right) \partial_{\varrho} + \frac{\bar{M}^2}{(1+\varrho^2)^2} -
\left( \frac{3}{4} \frac{\varrho^2}{(1+\varrho^2)^2} + \frac{1}{\varrho^2}(\ell+1)(\ell+3) - \frac{6}{1+\varrho^2} \right) \right] 
\psi^{\oplus}_{\calG_\ell}(\varrho) = 0. 
\end{equation}
The above admits analytic solutions -without any requirements as we did previously- which are given by: 
\begin{equation} \label{eq:nonormalizablesolution}
\begin{split}
\psi^{\oplus}_{\calG_\ell}(\varrho) = &(-)^{2 \ell} \frac{(1+\varrho^2)^{-\frac{1}{4} + \frac{\sqrt{1+\bar{M}^2}}{2}}}{\varrho^{\ell+3}} ~ 
{_2}F_1 \left(\frac{1}{2}+\frac{\sqrt{1+\bar{M}^2}}{2},-\frac{3}{2}-\ell+\frac{\sqrt{1+\bar{M}^2}}{2} ;-(\ell+1);-\varrho^2\right) \\
&+ \varrho^{\ell+1}(1+\varrho^2)^{-\frac{1}{4} + \frac{\sqrt{1+\bar{M}^2}}{2}}~{_2}F_1\left(\frac{1}{2}+\frac{\sqrt{1+\bar{M}^2}}{2},
 \frac{5}{2}+\ell+\frac{\sqrt{1+\bar{M}^2}}{2};\ell+3;-\varrho^2\right)\,.
\end{split}
\end{equation}
Substituting $\psi_0$ from \eqref{eq:simplemode} and $\psi_1$ from \eqref{eq:nonormalizablesolution} in the condition \eqref{eq:condition_2_schr}, 
and solving for $c$ we obtain an answer in terms of hypergeometric functions. When we fix $\ell=0$ the solution for $c$ simplifies to complex infinity. 

Similarly, one can ask the question of whether or not the ground state ($n=\ell=0$) of the setup can be driven to be massless or even tachyonic. Again, 
numerical analysis in \citep{Abt:2019tas} has shown that this does not happen as it requires a double-trace deformation with infinitely strong coupling. The 
same conclusion that we reached previously can also be drawn for this example. The only change in the computation is that now the $n=\ell=0$ state plays the
role of $\psi_1$ in \cref{eq:condition_2_schr} and the non-normalizable solution given by \cref{eq:nonormalizablesolution} plays the role of the state with a
zero mass. 

Consequently, we argue that we found an analytic way to see why a level crossing does not occur in these systems and furthermore why the discontinuity in 
the double-trace plots appeared to be going to infinite values of the coupling.
\section{The D4 background and  five dimensional SYM theories} \label{sec: D4}
For illustrative purposes we show explicitly the relevant brane intersections in the case of the D$4$ background, see \Cref{table: D4_brane_intersections}. 
\begin{table*}[h]
\begin{center}
\begin{tabular}{ |c|c|c|c|c|c|c|c|c|c|c|c|}
 \hline
 &&&&&&&&&&\\[-0.95em] 
   									& $x^0$ & $x^1$ & $x^2$ & $x^3$ & $x^4$ & $x^5$ & $x^6$ & $x^7$ & $x^8$ & $x^9$			\\ 
 \hline
 Background D$4$-brane 	& --- & --- & --- & --- & --- & $\bullet$ & $\bullet$ & $\bullet$ & $\bullet$ & $\bullet$		\\ 
 \hline
 Probe D$4$-brane 			& --- & --- & --- & $\bullet$ & $\bullet$ & --- & --- & $\bullet$ & $\bullet$ & $\bullet$				\\
 \hline
 Probe D$6$-brane 			& --- & --- & --- & --- & $\bullet$ & --- & --- & --- & $\bullet$ & $\bullet$				\\
 \hline
 Probe D$8$-brane 			& --- & --- & --- & --- & --- & --- & --- & --- & --- & $\bullet$												\\  
 \hline
\end{tabular}
\caption{The brane intersections that preserve $\mathcal{N}=2$ SUSY in a four-dimensional language. In the above notation --- denotes that a brane extends along that 
particular direction, while $\bullet$ means that the coordinate is transverse to the brane.}
\label{table: D4_brane_intersections}
\end{center}
\end{table*}

The fermionic action is given by \cite{Marolf:2003vf, Marolf:2003ye, Martucci:2005rb}
\begin{equation}\label{eq: D4_action}
S_{Dk} = \frac{T_{Dk}}{2} \int d^{k+1} \xi \sqrt{-\hat{g}}\, \bar{\Psi}\, \mathcal{P}_{-} \left[ \slashed{D}_{Dk} + \frac{e^{\phi} }{8 \cdot 4!} F_{ABCD} \left( 
\Gamma^{ABCD} - \Gamma^{\hat{M}} \Gamma^{ABCD} \Gamma_{\hat{M}} \right) - \frac{1}{2} \Gamma^M \partial_M \phi \right] \Psi\,.
\end{equation}

Varying \cref{eq: D4_action} with respect to the conjugate spinor $\bar\Psi$, we readily obtain the equation of motion of the spinor $\Psi$, 
namely it is
\eq$\label{eq: D4eqmot}
\slashed{D}_{Dk}\Psi + \frac{e^{\phi}}{8 \cdot 4!} F_{ABCD} \left( \Gamma^{ABCD} - \Gamma^{\hat{M}} \Gamma^{ABCD} \Gamma_{\hat{M}} \right)\Psi
-\frac{1}{2} (\Gamma^M \partial_M \phi) \Psi=0\,,$
where the dilaton field and the R-R field strength, which in this case constitutes a 4-form, can be determined from \cref{eq: dil&RRpot} with $p=4$, 
thus we have
\eq$\label{eq: dil&RR4}
e^{\phi}=\left(\frac{r}{R}\right)^{\frac{3}{4}}\,,\hspace{3cm} F_{(4)}=\star (dC_{(5)})\,,$
respectively. As it is depicted in \Cref{table: D4_brane_intersections}, there are three possibilities for the probe D$k$-branes, $k=\{4,6,8\}$ with 
$d=\{2, 4, 6\}$ respectively.  In case of the D$4$/D$4$-brane setup, the above relation leads to
\bal$\label{eq: RR44}
F_{(4)}=\frac{3}{r^2}&\left[\rho\, e^{(\theta)}\wedge e^{(w^1)}\wedge e^{(w^2)}\wedge e^{(w^3)}+e^{(\rho)}\wedge e^{(\theta)}\wedge\left(
w^1\, e^{(w^2)}\wedge e^{(w^3)}\right.\right.\nonum\\[2mm]
&\left.\left.-w^2\, e^{(w^1)}\wedge e^{(w^3)}+w^3\, e^{(w^1)}\wedge e^{(w^2)}\right)\right]\,.$
In case of the D$4$/D$6$-brane setup, it is
\bal$\label{eq: RR46}
F_{(4)}=\frac{3}{r^2}&\left[\rho\, e^{(S^2)}\wedge e^{(w^1)}\wedge e^{(w^2)}-e^{(\rho)}\wedge e^{(S^2)}\wedge\left(
w^1\, e^{(w^2)}-w^2\, e^{(w^1)}\right)\right]\,,$
where $e^{(S^{2})}=e^{(\theta_1)}\wedge e^{(\theta_2)}$. Finally, in case of the D$4$/D$8$-brane setup, we have
\bal$\label{eq: RR48}
F_{(4)}=\frac{3}{r^2}&\left(\rho\, e^{(S^3)}\wedge e^{(w^1)}+w^1\, e^{(\rho)}\wedge e^{(S^3)}\right)\,,$
where $e^{(S^3)}=e^{(\theta_1)}\wedge e^{(\theta_2)}\wedge e^{(\theta_3)}$. Having in our disposal the components of the R-R field strength as 
well as the expression of the dilaton field it is straightforward to calculate that for the D$4$/D$k$-brane system it is
\eq$\label{eq: gammaterm4}
\frac{e^{\phi}}{8 \cdot 4!}F_{ABCD}\left(\Gamma^{ABCD} - \Gamma^{\hat{M}} \Gamma^{ABCD} \Gamma_{\hat{M}}\right)\Psi=-
\frac{3}{4}\frac{\rho}{R^{3/4}\,r^{5/4}}\Gamma_{S^{k-d-1}w^1\cdots w^{5-k+d}}\Psi\,,$
while by setting $p=4$ in eqs. \eqref{eq: dirac&dilterm} we get
\bal$\label{eq: dirac&dil4}
\slashed{D}_{Dk}\Psi^{\ell^{\pm}}-\frac{\Gamma^M\pa_M\phi}{2}\, \Psi^{\ell^{\pm}}=&\left(\frac{R}{r}\right)^{\frac{3}{4}}\Gamma^\mu\pa_\mu
\Psi^{\ell^{\pm}}+\left(\frac{r}{R}\right)^{\frac{3}{4}}\Gamma^\rho\pa_\rho\Psi^{\ell^{\pm}} \pm\frac{1}{\rho}\left(\frac{r}{R}\right)^{\frac{3}{4}} 
\left(\ell + \frac{m}{2} \right)\Psi^{\ell^{\pm}}\nonum\\[2mm]
&+\frac{1}{R^{\frac{3}{4}}}\left(\frac{3}{8}\frac{\rho}{r^{\frac{5}{4}}}+\frac{m}{2}\frac{r^{\frac{3}{4}}}{\rho}\right)\Gamma^\rho\Psi^{\ell^{\pm}}\,.$
Substituting  eqs. \eqref{eq: gammaterm4}, \eqref{eq: dirac&dil4} in \eqref{eq: D4eqmot}  and imposing the projection  $\Gamma_{S^{k-d-1}w^1\cdots 
w^{5-k+d}}\Psi=-\Psi$, 
we are led to the following first-order equation of motion
\eq$\label{eq: first_order_D4}
\left\{ \left( \frac{R}{r} \right)^{\frac{3}{4}} \Gamma^{\mu} \partial_{\mu} +  \left( \frac{r}{R} \right)^{\frac{3}{4}} \Gamma^{\rho} \partial_{\rho} 
+\frac{1}{R^{\frac{3}{4}}} \left[\frac{3}{4} \frac{\rho}{r^{\frac{5}{4}}} \pm \left( \ell + \frac{m}{2} \right) \frac{r^{\frac{3}{4}}}{\rho} \right] +
\frac{1}{R^{\frac{3}{4}}}\left( \frac{3}{8} \frac{\rho}{r^{\frac{5}{4}}} + \frac{m}{2}\frac{r^{\frac{3}{4}}}{\rho} \right) \Gamma^{\rho}\right\}
\Psi^{\ell^{\pm}} = 0\,.$

Applying the procedure which was described in \cref{sec: general idea} to 
\cref{eq: first_order_D4}, we construct the following second-order differential equation
\eq$\label{eq: D4_sec_ord_eq}
\left( \mathcal{A}^{\pm}_{1} \partial_{\varrho}^2 + \mathcal{A}^{\pm}_{2} \partial_{\varrho} + \mathcal{A}^{\pm}_{3} \bar{M}^2 +
\mathcal{A}^{\pm}_{4} \Gamma^{\varrho} + \mathcal{A}^{\pm}_{5} \right) \psi^{\ell^{\pm}}(\varrho) = 0\,,$ 
where
\begin{equation} \label{D4: second_order_eom}
\begin{split}
\mathcal{A}^{\pm}_1 &= 1, \quad \mathcal{A}^{\pm}_2 = \frac{m}{\varrho} + \frac{9}{4} \frac{\varrho}{\varrho^2 +1}, \quad \mathcal{A}^{\pm}_3 = 
\frac{1}{\left(\varrho^2 +1 \right)^{3/2}}\,, \\
\mathcal{A}^{\pm}_4 &= \frac{3}{4} \frac{1\pm 2 \ell \pm m}{\varrho^2+1} \mp \frac{1}{\varrho^2} \left( \ell + \frac{m}{2} \right) - \frac{3}{8} 
\frac{\varrho^2}{\left( \varrho^2 + 1 \right)^2}\,, \\
\mathcal{A}^{\pm}_5 &= \frac{3}{8} \frac{1}{\varrho^2+1} \left( 3m \mp 2m + 1 \mp 4 \ell \right) - \frac{39}{64} \frac{\varrho^2}{(\varrho^2+1)^2}
 - \frac{1}{\varrho^2} \left( \ell^2 + m \ell+\frac{m}{2} \right)\,.
\end{split}
\end{equation}
Employing the projection $\Gamma^\rho\, \psi^{\ell^{\pm}}=\psi^{\ell^{\pm}}$ and using eqs. \eqref{eq: D4_sec_ord_eq} and 
\eqref{D4: second_order_eom} one can readily obtain the fermionic second-order ordinary differential equation for $\psi^{\ell^{\pm}}(\varrho)$. 
In case of positive spinorial harmonics ($\ell^+$), we get the differential equation of $\psi^{\oplus}_{\calG_{\ell}}(\varrho)$, while for negative 
spinorial harmonics ($\ell^-$), the differential equation of $\psi^{\oplus}_{\calF_{\ell}}(\varrho)$ is obtained. The corresponding bosonic equations of 
motion for the D$4$/D$k$-brane setups ($k=\{4,6,8\}$), can be determined from eqs. \eqref{eq: bosonic_spectra_4}-\eqref{eq: bosonic_spectra_0} by 
setting $p=4$. For all possible D$4$/D$k$-brane systems the transformation 
\eq$\label{eq: pos_spher_trans_4}
\psi^{\oplus}_{\calG_{\ell}}(\varrho)=\frac{1}{(1+\varrho^2)^{9/16}}\, f_{\ell+1}(\varrho)\,,$ 
maps the fermionic function $\psi^{\oplus}_{\calG_{\ell}}(\varrho)$ to the bosonic function $f_{\ell+1}(\varrho)$,  while the transformation
\eq$\label{eq: neg_spher_trans_4}
\psi^{\oplus}_{\calF_{\ell}}(\varrho)=\frac{1}{(1+\varrho^2)^{9/16}}\, f_{\ell}(\varrho)\,,$
maps the fermionic function $\psi^{\oplus}_{\calF_{\ell}}(\varrho)$ to the bosonic function $f_{\ell}(\varrho)$. Notice as well that the above 
transformations can be obtained from \eqref{eq: basic_1} by setting $p=4$. 

\begin{table}[H]
    \centering
   \begin{tabular}{|c|c|c|c|c|c|} 
   \cline{3-6}
    \multicolumn{2}{c}{}     & \multicolumn{4}{|c|}{$\boldsymbol{n}$} \\ \cline{3-6}
    \multicolumn{2}{c|}{}        & 0 & 1 & 2 & 3       \\ \cline{1-6}
    \multirow{4}{0.5em}{$\boldsymbol{\ell}$}  
    						  & 0 & 	5.0604 & 14.068 & 27.856 & 46.573 \\ \cline{2-6}  
                             & 1 &	14.685 & 26.910 & 43.421 & 64.677 \\ \cline{2-6}
                             & 2 &	29.701 & 45.647 & 65.237 & 89.117 \\ \cline{2-6}
                             & 3 & 49.985 & 69.994 	& 93.198 & 120.08 \\ \cline{1-6}
  \end{tabular}
\hfill%
    \begin{tabular}{|c|c|c|c|c|c|} 
    \cline{3-6}
    \multicolumn{2}{c}{}     & \multicolumn{4}{|c|}{$\boldsymbol{n}$} \\ \cline{3-6}
    \multicolumn{2}{c|}{}        & 0 & 1 & 2 & 3       \\ \cline{1-6}
    \multirow{4}{0.5em}{$\boldsymbol{\ell}$}  
    						  & 0 & 	2.2741 	& 9.8420 & 22.321 & 39.768 \\ \cline{2-6}  
                             & 1 &	9.1908 & 19.745 & 34.871 & 54.857 \\ \cline{2-6}
                             & 2 &	21.527 & 35.556 & 53.543 & 76.084 \\ \cline{2-6}
                             & 3 &	39.189 & 57.140 & 78.472 & 103.79 \\ \cline{1-6}
  \end{tabular}
\newline
\vspace*{1 cm}
\newline
    \begin{tabular}{|c|c|c|c|c|c|} 
    \cline{3-6}
    \multicolumn{2}{c}{}     & \multicolumn{4}{|c|}{$\boldsymbol{n}$} \\ \cline{3-6}
    \multicolumn{2}{c|}{}        & 0 & 1 & 2 & 3       \\ \cline{1-6}
    \multirow{4}{0.5em}{$\boldsymbol{\ell}$}  
    						  & 1 & 	5.0604 & 14.068 & 27.856 & 46.573 \\ \cline{2-6}  
                             & 2 &	14.685 & 26.910 & 43.422 & 64.677 \\ \cline{2-6}
                             & 3 &	29.701 & 45.647 & 65.237 & 89.116 \\ \cline{1-6}
  \end{tabular}
\caption{Numerical results of the parameter $\bar{M}^2$ for the D$4$/D$k$ brane intersections. The top left table pertains to the D$4$/D$8$ 
and the top right one depicts results for the D$4$/D$6$ brane intersection. The bottom table is related to the D$4$/D$4$ setup.}
\label{tab: d4}
\end{table}

The above equations of motion cannot be solved analytically (as in the case of bosonic mesons) and we have to resort to numerical analysis. We show 
explicitly the first few numerical values of the masses in \Cref{tab: d4}. The numerical values for $n$ and $\ell$ describe the $\mathcal{F}$ operators and 
one needs to shift appropriately the $\ell$ to read off the value of the $\mathcal{G}$ states.
\section{Epilogue} \label{sec: outro}
In this work we have considered BPS brane intersections in Type IIA/B theories in static equilibrium that include fields in the fundamental representation of 
the gauge group. We have studied and derived systematically the equations of motion from the fermionic completion of the DBI action. We have showed the
degeneration between the bosonic and fermionic states by providing the necessary transformations to map the equations of motion from the latter to the 
former ones. We state the transformation rules once more. The map is given by: 
\begin{equation*}
\psi^{\oplus}_{\mathcal{G}_\ell}(\varrho) = \frac{1}{(1+\varrho^2)^{\frac{3}{16}(7-p)}} f_{\ell+1}(\varrho), \qquad 
\psi^{\oplus}_{\mathcal{F}_\ell}(\varrho) = \frac{1}{(1+\varrho^2)^{\frac{3}{16}(7-p)}} f_{\ell}(\varrho),
\end{equation*}
where in the above we denote by $\oplus$ superscript the supergravity modes obtained by considering the projection $\Gamma^\rho\Psi^{\oplus}=\Psi^{\oplus}$ 
and we reserve the $\ominus$ for the negative eigenvalue of the projection. 

In addition to that, we have checked the supersymmetric degeneracy of the bosonic and fermionic masses numerically in each probe-brane system. We have 
explicitly computed the value of $\bar{M}^2$  by solving the bosonic as well as the fermionic equations of motion and derived the same masses as expected. We 
remind at this stage and for the final time, that we have already performed the appropriate shifts in the $\ell$ quantum number as described in 
\cite{Kruczenski:2003be} in order to keep the presentation of our tables brief. We have included our numerical estimations for the masses in the various sections 
of the main body of the paper, see \Crefrange{tab: d0}{tab: d4}. Hence, we have derived the necessary equations to study the dynamics of world-volume fermions in holographic top-down systems and performed crucial checks regarding their validity.

As a by-product of our studies we managed to show that these states obey the field theory mass scaling relation that baryons have in the large-$N_c$ limit when we 
fix the other quantum numbers ($n$ which counts the nodes of the wavefunction and $\ell$ which is the angular momentum of the fields)
\begin{equation*}
M^2 \sim N^2_c.
\end{equation*}
We also discussed what we consider to be the precise interpretation of this result. Namely we concluded that this result should be valid if one manages to solve the 
true dynamic string baryon vertex in the limit where the field theory is at the conformal window (exhibits walking dynamics). 

We would like to comment on further new possible directions below: 
The most straightforward generalisation of our considerations can be performed in the context of eleven-dimensional supergravity. The supersymmetric M-brane 
junctions are M$2$/probe-M$5$, M$5$/probe-M$5$ and M$2$/probe-M$2$. The bosonic cases have been performed already in \cite{Arean:2006pk} where it 
was found that the equations of motion coincide with the D$1$/probe-D$5$, D$4$/probe-D$4$ and the F$1$/probe-D$2$ systems respectively. In this notation 
by F$1$ we denote the fundamental string. We believe that the approach we developed here is directly applicable in these systems.

One of the approaches that has been followed in order to construct gravity duals of more realistic gauge theories with a reduced amount of symmetry is to replace 
the five-dimensional sphere of the $AdS_5 \times S^5$ background with a five-dimensional Sasaki-Einstein space. We denote the latter generically by 
$\mathcal{M}^5$. Doing so we obtain a duality between string theory in $AdS_5 \times \mathcal{M}^5$ and a super-conformal quiver theory. By now we know many 
such models explicitly. The first example that was studied in the literature is the so-called Klebanov-Witten model where we have $\mathcal{M}^5 = T^{1,1}$ 
\cite{Klebanov:1998hh}. We also have at our disposal infinite five (and also higher) dimensional spaces of cohomogeneity $1$ and $2$. These manifolds are called 
$Y^{p,q}$ \cite{Gauntlett:2004yd} and $L^{p,q,r}$ \cite{Cvetic:2005ft, Martelli:2005wy} respectively, with already existing work on supersymmetric brane 
embeddings in these backgrounds \cite{Canoura:2005uz, Canoura:2006es}. It would be an interesting project to derive the corresponding mass spectra of mesonic 
states in these cases. 

It would also be interesting to derive the fermionic spectra of gravity duals that describe chiral symmetry breaking ($\chi$SB). The most straightforward example 
of $\chi$SB in the context of holography is provided by the D$3$/probe-D$7$ with a non-trivial NS flux (Kalb-Ramond field) on the probe-brane 
\cite{Erdmenger:2007bn}. The main conceptual difference between this system of $\chi$SB and the systems that we have considered here, is the non-trivial dependence of the embedding function on the holographic radial coordinate. In the magnetic field example instead of having the separation of the background and the probe branes to be a constant, it is a non-trivial function $L(\rho)$ that depends on the holographic radial coordinate. Geometrically this means that the probe-brane is curved. The practical issue in cases with such embeddings is the manipulation 
of the $\Gamma$-matrix projections in order to construct an ordinary second-order differential equation to apply the general steps we developed here. 

Finally, it would be interesting to understand the dynamics of world-volume fermions in anisotropic RG flows in type IIB supergravity. These RG flows are solutions  from the isotropic fixed points to the anisotropic ones. There are solutions in the literature \cite{Azeyanagi:2009pr} that describe the D$3$/D$7$ brane setup which extend from the familiar $AdS_5$ solution in the UV to the anisotropic solution in the IR. This generalisation of the AdS/CFT duality to include the description of Lifshitz-like fixed points was initiated in \cite{Kachru:2008yh} and is interesting in the context of condensed matter theory and applications thereof. We believe that the general approach we developed here is applicable in these systems as well, and might also make computations more feasible. 
\section*{Acknowledgements}
We would like to thank Nick J. Evans and Johanna Erdmenger for collaboration on related topics and discussions, as well as Daniel Ar\'{e}an and Alfonso V.~Ramallo for sharing their private notes at the initial stages of this project. We are thankful to Nick J. Evans and Panagiota Kanti for a crucial read and comments on the final draft of this paper. The research of T.N. was co-financed by Greece and the European Union (European Social Fund- ESF) through the Operational Programme “Human Resources Development, Education and Lifelong Learning” in the context of the project “Strengthening Human Resources Research Potential via Doctorate Research – 2nd Cycle” (MIS-5000432), implemented by the State Scholarships Foundation (IKY). Both authors would like to dedicate this work to the memory of Pavlos Fyssas. 
\setcounter{equation}{0}
\appendix
\newpage

\section{Conventions and notation} \label{app: conventions}
We find it necessary to collect and clarify our conventions and notation for the various manipulations of the main body of our work in this appendix. We 
begin by discussing the different letters that have appeared throughout the various sections. 

Capital letters $A,B,C,$... denote type IIA/B coordinates and they take values in the range $(0, \ldots, 9)$. The space described by the $Z$-coordinates 
corresponds to spatial coordinates which are transverse to the background D$p$-branes. Its dimensionality is equal to $\left( 9-p \right)$. 
Lower-case Greek letters $\mu, \nu,$... are the Minkowski indices which correspond to the spacetime coordinates of the D$p$-brane . The letters $\rho$ 
and $\varrho$ denote the radial and the rescaled dimensionless  radial coordinate, respectively. Hatted indices $\hat{A}, \hat{B},$...
parametrize the spacetime spanned by the probe D$k$-brane. $Y$-coordinates denote the ($k-d$)-dimensional internal space to the probe brane. Finally, 
we use the set $\{w^1, w^2, \ldots, w^{9-p-k+d} \}$ of coordinates to specify the subspace of the original ten-dimensional geometry that is transverse to both the
background and the probe branes. Indices enclosed within brackets, i.e $(A), (\mu)$, take values according the aforementioned rules and denote the passing to the 
vielbein basis.

For the components of a $p$-form we use the convention
\eq$
\mathcal{X}_{(p)} \equiv \frac{1}{p!} ~ \mathcal{X}_{\mu_1 \mu_2\ldots\mu_p} ~ dx^{\mu_1} \wedge dx^{\mu_2} \wedge \ldots \wedge dx^{\mu_p}=
\mathcal{X}_{|\mu_1 \mu_2\ldots\mu_p|} ~ dx^{\mu_1} \wedge dx^{\mu_2} \wedge \ldots \wedge dx^{\mu_p}.$
In the above relation, the vertical bars denote that only components with increasing indices are included in the summation.

\textbf{Definition of the Hodge star operator:} Let us consider an $n$-dimensional spacetime and the usual $\{dx^M\}$ basis which satisfies the
relation $ds^2=g_{\mu\nu} dx^\mu dx^\nu$, then the \textit{dual} of the $q$-form (with $q<n$) $dx^{\mu_1}\wedge \ldots\wedge dx^{\mu_q}$ is defined as
\bal$\label{eq: dual-basis}
\star \left(dx^{\mu_1}\wedge dx^{\mu_2}\wedge \ldots\wedge dx^{\mu_q}\right)&\equiv\frac{\sqrt{-g^{(n)}}}{(n-q)!}\,g^{\mu_1 \nu_1}\,g^{\mu_2 \nu_2}
\cdots g^{\mu_q \nu_q}\,\epsilon_{\nu_1 \nu_2\ldots \nu_q\,\nu_{q+1}\ldots \nu_n}\, dx^{\nu_{q+1}}\wedge\ldots\wedge dx^{\nu_n}\nonum\\[2mm]
&=\sqrt{-g^{(n)}}\,g^{\mu_1 \nu_1}\,g^{\mu_2 \nu_2}\cdots g^{\mu_q \nu_q}\,\epsilon_{\nu_1 \nu_2\ldots \nu_q\,|\nu_{q+1}\ldots \nu_n|}\, 
dx^{\nu_{q+1}}\wedge\ldots\wedge dx^{\nu_n}\,,$
where $g^{(n)}=\text{det}(g_{\mu\nu}) $, $\epsilon_{012\ldots n}=1$. The star $\star$ is called \textit{Hodge star operator}. Using the above equation, we can determine the dual of a 
$p$-form $\calX_{(p)}$, namely it is
\eq$\label{eq: dual-form}
\star \calX_{(p)}=\sqrt{-g^{(n)}}\,\calX_{|\mu_1\mu_2\ldots\mu_p|}g^{\mu_1 \nu_1}\,g^{\mu_2 \nu_2}\cdots g^{\mu_p \nu_p}\,
\epsilon_{\nu_1 \nu_2\ldots \nu_p\,|\nu_{p+1}\ldots \nu_n|}\, dx^{\nu_{p+1}}\wedge\ldots\wedge dx^{\nu_n}\,,$
with $p<n$.

Let us now consider the vielbein basis $\{e^{(\mu)}\}$, which satisfies the relation $ds^2=\eta_{(\mu)(\nu)}e^{(\mu)}e^{(\nu)}$. In this case, we
would have
\bal$\label{eq: dual-vielbein}
\star \left(e^{(\mu_1)}\wedge \ldots\wedge e^{(\mu_q)}\right)&\equiv \eta^{(\mu_1)(\nu_1)}\cdots
\eta^{(\mu_q)(\nu_q)}\,\epsilon_{(\nu_1)\ldots (\nu_q)\,|(\nu_{q+1})\ldots (\nu_n)|}\,e^{(\nu_{q+1})}\wedge\ldots\wedge e^{(\nu_n)}\,.$

\newpage
\section{Vielbeins and spin-connection components of a unit N-sphere} \label{app: unitsphere}
The line-element which defines the geometry of a unit $N$-sphere is of the following form
\eq$ds^2=d\Omega_N^2=d\theta_1^2+\sum_{i=2}^{N}\left(\prod_{j=1}^{i-1}\sin^2\theta_j\right)d\theta_i^2\,.$
It is straightforward to deduce that the vielbeins for the geometry above are given by
\eq$
e^{(\bar{m})}{}_{\bar{1}}=\delta^{\bar{m}}{}_{\bar{1}}, \hspace{2em}e^{(\bar{m})}{}_{\bar{n}}=\displaystyle{\prod_{j=1}^{n-1}\sin\theta_j\ 
\delta^{\bar{m}}{}_{\bar{n}}},\hspace{1em}  1<n\leq N,$
while the spin-connection components can be expressed in a compact form as
\bal$ 
\omega^{(\bar{m})(\bar{n})}_{\, \bar{i}}=&\sum_{k=2}^N \delta^{\bar{k}}{}_{\bar{i}}\sum_{j=1}^{k-1}\delta^{\bar{m}}{}_{\bar{k}}\,
\delta^{\bar{n}}{}_{\bar{j}}\cot\theta_j \prod_{\ell=j}^{k-1}\sin\theta_\ell-\sum_{k=2}^N\delta^{\bar{k}}{}_{\bar{i}}\,
\delta^{\bar{m}}{}_{\overline{k-1}}\,\delta^{\bar{n}}{}_{\bar{k}}\cos\theta_{k-1}\nonum \\[2mm]
&-\sum_{k=3}^{N}\delta^{\bar{k}}{}_{\bar{i}}\sum_{j=1}^{k-2}\delta^{\bar{m}}{}_{\bar{j}}\,\delta^{\bar{n}}{}_{\bar{k}}
\cos\theta_j\prod_{\ell=j+1}^{k-1}\sin\theta_\ell\,.$
In the above expressions we kept the same bar notation to be consistent with \cref{sec: generalformulae}.
\section{Scalar mesons, probe branes and numerical spectra} \label{app: numerical}
There has been extensive work in the literature studying the bosonic spectra and states of probe-brane setups. Here we will limit ourselves to a minimal discussion and 
quoting the basic relations. This we do for convenience with our numerical approach. Original and detailed work on this can be found in \cite{Myers:2006qr, Arean:2006pk}. 

Firstly, it is more useful to re-organize the probe brane junction in the following way: we have D$p$/D$p+4$, D$p$/D$p+2$, and D$p$/D$p$ systems. In the first class 
of probe-brane embeddings, $p$ is allowed to have values $0 \leq p \leq 4$. For co-dimension one defect $p$ can be $1 \leq p \leq 4$. Finally, for D$p$/D$p$ systems 
we have $2 \leq p \leq 4$. 

In all of the above systems, when considering an appropriate shift in the $\ell$ quantum number, the dynamics of the bosonic degrees of freedom can be mapped to those 
of scalar orthogonal fluctuations from the DBI action of the form 
\begin{equation}
Y^A = \delta^A_9 ~ L + 2 \pi {\alpha}^\prime \varphi^A 
\end{equation} 
with $A$ ranging from $5, \cdots, 9-p$, $4, \cdots, 9-p$ and $3, \cdots, 9-p$ for the  D$p$/D$p+4$, D$p$/D$p+2$, and D$p$/D$p$ systems respectively. 
Therefore, the most basic equations of motion that needs to be solve in the one pertaining to the above modes and we have: 
\bal$\label{eq: bosonic_spectra_4}
\partial^2_{\varrho} f_{\ell}(\varrho) + \frac{3}{\varrho} \partial_{\varrho} f_{\ell}(\varrho) + \left( \frac{\bar{M}^2}{(1 + \varrho^2)^{(7-p)/2}} - 
\frac{\ell(\ell+2)}{\varrho^2} \right) f_{\ell}(\varrho) &= 0, \qquad \textsl{for the D$p$/D$p+4$\,,} \\[2mm]
\label{eq: bosonic_spectra_2}
\partial^2_{\varrho} f_{\ell}(\varrho) + \frac{2}{\varrho} \partial_{\varrho} f_{\ell}(\varrho) + \left( \frac{\bar{M}^2}{(1 + \varrho^2)^{(7-p)/2}} - 
\frac{\ell(\ell+1)}{\varrho^2} \right) f_{\ell}(\varrho) &= 0, \qquad \textsl{for the D$p$/D$p+2$\,,} \\[2mm]
\label{eq: bosonic_spectra_0}
\partial^2_{\varrho} f_{\ell}(\varrho) + \frac{1}{\varrho} \partial_{\varrho} f_{\ell}(\varrho) + \left( \frac{\bar{M}^2}{(1 + \varrho^2)^{(7-p)/2}} - 
\frac{\ell^2}{\varrho^2} \right) f _{\ell}(\varrho) &= 0, \qquad \textsl{for the D$p$/D$p$}\,,$
where in the above we have considered an appropriate decomposition of the $ \varphi^A $ and thus $f$ is a scalar function of the radial coordinate. The final equation 
does not have normalizable solutions for zero angular momentum excitation and therefore the states with $\ell=0$ are considered non-physical. 

In \cite{Myers:2006qr, Arean:2006pk} the authors have given explicitly tables for some cases. For direct comparison, convenience and completeness, here we generate 
the numerical results and present them in tables that are complementary to those ones. 

Our numerical approach for the bosonic spectra is the following: 
\begin{itemize}
\item We are shooting from the $\Lambda_{\rm{IR}}$ to the $\Lambda_{\rm{UV}}$ by fine-tuning $\bar{M}^2$ such that the mode solutions are normalizable in the UV 
and small in amplitude \cite{Kruczenski:2003be}. 
\item We choose as initial conditions $f(\varrho)|_{\varrho \rightarrow 0} = \varrho^{\ell}$ and $\partial_{\varrho} f(\varrho)|_{\varrho \rightarrow 0} = \ell 
\varrho^{\ell-1}$ \cite{Myers:2006qr, Arean:2006pk}.
\item We use  $\Lambda_{\rm{IR}} = 10^{-7}$ and  $\Lambda_{\rm{UV}} = 10$.
\end{itemize}

\newpage

\clearpage
\bibliography{lit}{}
\bibliographystyle{utphys}

\end{document}